\documentclass[lettersize,journal]{IEEEtran}

\IEEEoverridecommandlockouts
\usepackage{cite}
\usepackage{amsmath,amssymb,amsfonts}
\usepackage{graphicx,color}
\usepackage{textcomp}
\usepackage{algorithm}
\usepackage{algorithmicx}
\usepackage{algpseudocode}
\usepackage{setspace}
\usepackage{bm}
\usepackage{xcolor}
\usepackage{tikz}
\usetikzlibrary{quantikz2}
\usepackage{float}
\usepackage{subcaption}
\usepackage[T1]{fontenc}
\usepackage{verbatim}
\usepackage{hyperref}
\usepackage{makecell}
\usepackage{booktabs}
\usepackage{array} 
\usepackage{amsthm}
\usepackage{graphbox} 
\usepackage{adjustbox}
\usepackage{balance}
\usepackage{multirow}
\usepackage[capitalize]{cleveref}
\usepackage{footnote}
\usepackage{bm}
\usepackage{threeparttable}
\usepackage{anyfontsize}

\newtheorem{theo}{Theorem}
\newtheorem{lem}{Lemma}
\newtheorem{cor}{Corollary}
\theoremstyle{remark}

\theoremstyle{definition}
\newtheorem{defin}{Definition}

\newtheorem*{remark}{Remark}

\newtheorem*{prob}{Research Problem}

\newcommand{\eqdef}{\stackrel{\triangle}{=}}

\def\BibTeX{{\rm B\kern-.05em{\sc i\kern-.025em b}\kern-.08em
    T\kern-.1667em\lower.7ex\hbox{E}\kern-.125emX}}

\usepackage{balance}

\begin{document}

\title{On the Efficient Extraction of  Entangled Resources}

\author{Si-Yi~Chen, Angela~Sara~Cacciapuoti and Marcello~Caleffi
    \thanks{A preliminary conference version of this work is \cite{CheCacCal-25-QCE}.\\ 
    The authors are with the \href{www.quantuminternet.it}{Quantum Internet Research}  Group, University of Naples Federico II, Naples, 80125 Italy.}
    \thanks{Corresponding author: Angela Sara Cacciapuoti (e-mail: angelasara.cacciapuoti@unina.it).}
    \thanks{This work has been funded by the European Union under Horizon Europe ERC-CoG grant QNattyNet, n.101169850. Views and opinions expressed are however those of the author(s) only and do not necessarily reflect those of the European Union or the European Research Council Executive Agency. Neither the European Union nor the granting authority can be held responsible for them.}
    }


\maketitle

\begin{abstract}
In the Quantum Internet, multipartite entanglement enables a rich and dynamic overlay topology, referred to as artificial topology,
upon the physical one, that can be exploited for communication purposes. In fact, the ability to extract $n$-qubits GHZ states and EPR pairs from the original multipartite entangled state constitutes the resource primitives for end-to-end and on-demand quantum communications. Thus, in this paper, we theoretically determine upper and lower bounds for the number of extractable $n$-qubits GHZ states and EPR pairs involving nodes remote in the artificial topology, as well as the achievable size $n$ of remote GHZ states. The theoretical analysis is then complemented by the proposal of a novel algorithm, which provides in
polynomial-time a heuristic solution to the above problem. This is remarkable, since the theoretical problem is NP-complete.  The performance analysis demonstrates the proposed algorithm is able to effectively manipulate the original and arbitrary graph state for extracting entanglement resources across remote nodes.
\end{abstract}


\maketitle

\section{Introduction}
\label{sec:1}

\renewcommand{\arraystretch}{1.5}
\begin{table*}
    \centering
    \caption{Adopted terms in entanglement-enabled connectivity domain}
    \label{tab:00}
    \fontsize{8pt}{8pt}\selectfont
    \begin{subtable}[t]{\textwidth}
    \centering
        
        {\begin{tabular}[t]{  p{3.5cm} | p{13cm}  }
        \hline
        \hline 
        \multicolumn{1}{c}{Terms} & {Interpretations} \\
        \hline
            \textit{Artificial topology} & A virtual network topology, built upon the physical topology, and associated with a certain multipartite entanglement state.\\
            
            \textit{Artificial link}
            & A virtual link, pictorially visualized as an edge between two nodes connected in an artificial topology, corresponds to a CZ interaction between the qubits and denotes the ``possibility'' of extracting an EPR pair between the two considered nodes. Thus, artificial link and EPR are not synonymous.\\

            \textit{Artificial subnet} & A virtual subnet, pictorially visualized  as a fully connected subgraph in an artificial topology, corresponds to the ``possibility'' of extracting a GHZ state among the involved nodes. \\

            \textit{(Artificial) remote nodes} & Non-adjacent nodes in the artificial topology that are not directly connected by an artificial link (Def.~\ref{def:01}).\\

            \textit{(Artificial) remote subnet}  & An artificial subnet formed by an independent set of nodes in an artificial topology.\\
            
            \textit{Ultimate artificial links}
            & The actual EPR pairs extracted from the original multipartite state.\\

            \textit{Ultimate artificial subnets} & The actual GHZ states extracted from the original multipartite state.\\

            \textit{Location} & The location of an (ultimate) artificial link/subnet refers to the identities of the interconnected nodes.\\

            \textit{Volume} & The volume of (ultimate) artificial link/subnet refers to the number of EPR pairs / GHZ states that can be \textit{simultaneously} extracted from a given multipartite state. This volume heavily depends on the type and structure of the considered multipartite state, and some of the artificial links are depleted during the extraction process.\\

            \textit{Mass} & The mass of an (ultimate) artificial subnet refers to the number of interconnected nodes.\\

        \hline   
        \end{tabular}}
    \end{subtable}
\end{table*}

\IEEEPARstart{E}{ntanglement} shared between more than two parties, known as multipartite entanglement, represents a powerful resource for quantum networks \cite{CalCac-25,CacIllCal-23,PirDur-18,PirDur-19,IllCalMan-22,RamPirDur-21,KruAndDur-04,IllCalVis-23,LiuLiCai-24,LiuLiWang-25}. It enables a new form of connectivity, referred to as \textit{entanglement-enabled connectivity} \cite{IllCalMan-22,CacIllCal-23}, which augments the physical topology with virtual links, activated by the entanglement, and referred to as \textit{artificial links}, 
between pairs of nodes, remote in the physical topology\footnote{It is worth noting that, in agreement with current quantum technology Technology Readiness Level (TRL), the physical network topology is generally sparse. Thus, it heavily limits the node communication capabilities~\cite{MazCalCac-24-QCNC}.}, without any additional physical link deployment. Thus, multipartite entanglement enables a richer, dynamic overlay topology, referred to as \textit{artificial topology}, upon the physical one. 
And this 
artificial topology can be properly manipulated to account for the dynamics of the node communication needs \cite{CheIllCac-24,MazCalCac-24,CheIllCac-24-QCE}. 

Most of the literature on multipartite entanglement manipulation usually aims at extracting from the initial multipartite state a certain amount, say $k$, of shared EPR pairs. These $k$ EPR pairs can be subsequently exploited for the parallel “transmission” of $k$ informational qubits, by adopting the quantum teleportation protocol \cite{CacCalVan-20}. It is worthwhile to note that the identities of the nodes involved in the $k$ disjoint pairs are fixed, with no possibility of changing them to time-varying communication needs. 
Differently, the manipulation of an artificial topology for extracting GHZ states \cite{ManPat-23,BriRau-01,FraLiuRai-23,JonHahTch-24} overcomes the above constraint. Specifically, a GHZ state -- key for various quantum communication protocols~\cite{VanEmiGag-24,ChrWeh-05,MarSan-11,MurGraKam-20,HahJonPap-20} -- represents, from a communication perspective, an artificial subnet, extracted among a certain number of nodes, starting from the original multipartite state. And the rationale for defining a GHZ as subnet rather than link is that it enables the dynamic extraction of an EPR pair between any pair of nodes sharing the original GHZ state. Remarkably, this extraction can happen at run-time, depending on the actual node communication needs.  From the above it follows that nodes belonging to an artificial subnet exhibit an \textit{entanglement proximity}, i.e. a distance in terms of entanglement hops, equal to one.

In this context, it is key to observe that having a fully-connected artificial topology among all the network nodes is not reasonable, due the challenges related to (and the complex equipment necessary for) the generation and the control of a high-order multipartite state. Hence, it is more practical and reasonable to assume the presence
of nodes that are remote even in the artificial topology. In the remaining part of this work, we focus on this type of nodes, since remote nodes in the artificial topology face reduced communication opportunities compared to nodes already interconnected. As a consequence, artificially interconnecting this type of nodes, not only assures network fairness, but also constitutes a communication primitive for end-to-end and on-demand communications \cite{PomHerBai-21,HerPomBeu-22,LiLiLiu-21,HumKalMor-18,BenHajVan-24}.    

Additionally, it is key to observe that the entanglement extraction capabilities of a certain multipartite state heavily depend on the features of this selected state \cite{HeiDurEis-06}. The original state also affects whether the extractions happen deterministic or probabilistic \cite{IllCalMan-22}. Thus, the choice of the initial multipartite state is a key network design choice. 
A notable class of multipartite entangled states is the \textit{two-colorable graph state} class \cite{Bri-05}, modeling important communication network topologies, such as grid, star, bistar, linear, even loop, butterfly, cluster networks \cite{TanZhaKra-23,CheIllCac-24,FreDur-24,HahPapEis-19,BraShaSze-22,MazCalCac-24,BenHajVan-24,AzuTamLo-15,MazZhaChu-25}.

With the above in mind, in this paper we assess the extraction capability of a generic graph state, in terms of ``volume'', ``mass'' and ``location''\footnote{We collected and summarized the terms widely exploited in the remaining part of the manuscript in Tab.~\ref{tab:00}.} of GHZ states and EPR pairs, shared among nodes \textit{remote in the original artificial topology}. And we refer to the aforementioned  remote extraction capability for GHZ and EPR states as \textit{remote Gability} and \textit{remote Pairability}, respectively. 
In a nutshell, we: 
\begin{itemize}
    \item \textit{quantify the volume} of ultimate links via lower bounds, by assessing the extraction EPR capabilities starting from an arbitrary two-colorable graph state;
    \item \textit{quantify both volume and mass of ultimate subnets} via lower bounds, by assessing the $n$-qubit GHZ extraction capabilities for any size $n$;
    \item \textit{quantify} the remote Pairability and $n$-Gability volumes also via theoretical upper bounds.
\end{itemize}

%

%
The theoretical analysis is then complemented by the proposal of a novel algorithm, which provides in polynomial-time a heuristic solution to the above problem. This is remarkable, since the theoretical problem is NP-complete, as better highlighted in the following subsection. 

The rest of this manuscript is organized as follows. In Sec.~\ref{sec:2} we first provide the reader with a formal definition of the research problem, along with an overview of the main results derived in the manuscript. In Sec.~\ref{sec:3}, we first present some preliminaries, and then we derive constructive conditions for both the remote Gability and remote Pairability. In Sec.~\ref{sec:4}, we present the proposed polynomial-time algorithm along with its complexity analysis. In Sec.~\ref{sec:5}, we evaluate the tightness of the constructive derived bounds with respect to general and bipartite graph states.  Finally, in Sec.~\ref{sec:6} we conclude the paper and we discuss key aspects related to the proposed approach.

\subsection{Related Work And Contribution}

\begin{figure}[t]
    \centering
    \resizebox{0.5\textwidth}{!}{
        \input{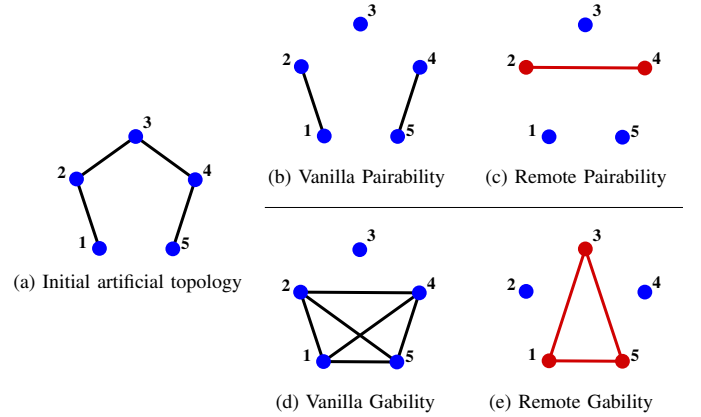}
    }
    \caption{\textit{Remote} vs \textit{Vanilla} Pairability and Gability for a 5-qubit linear graph state. (a) The initial artificial topology is a 5-qubit linear graph state. (b) Vanilla Pairability allows extraction of up to two EPR pairs from (a). (c) Remote Pairability enables extraction of only one EPR pair between remote nodes from (a). (d) Vanilla Gability permits extraction of a maximal 4-qubit GHZ state from (a). (e) Remote Gability supports extraction of a maximal 3-qubit GHZ state among remote nodes, corresponding to the maximum independent set in (a).}
    \label{fig:01}
    \hrulefill
\end{figure}

In \cite{DahHelWeh-18}, it has been shown that the computational complexity of extracting Bell pairs from graph states is NP-complete. And in the same paper, the authors refer to this problem as Bell-VM \cite{DahHelWeh-18}.
Research in this area~\cite{BraShaSze-22,CauClaMha-24,FreDur-24,DahHelWeh-18,HahPapEis-19} has potential applications in point-to-point quantum communication protocols, where the extracted $k$ Bell pairs can be used for the parallel transmission of $k$ informational qubits via teleporting protocol. In particular, in \cite{BraShaSze-22}, it is showed that $N$-qubit CSS state can extract $k$ pairs of EPR between $k$ disjoint parties, with $k$ proportional to $\log N$. Extending this idea, \cite{CauClaMha-24} determines two families of $k$-vertex-minor universal graphs based on two-colorable graph states, starting from a $N=O(k^4)$-qubit resource state. Similarly, \cite{FreDur-24} proposes the Zipper-protocol, enabling the extraction of multiple EPR pairs from a 2D cluster state along the diagonal direction. In contrast, the X-protocol in \cite{HahPapEis-19} extracts EPR pairs at predetermined locations while somehow preserving part of the entanglement among remaining qubits.

A related line of research focuses on determining whether a GHZ state can be extracted from a given graph state, by using only local Clifford (LC) operations, local Pauli measurements (LPM), and classical communication (CC). This problem has also been proven to be NP-complete~\cite{DahHelWeh-20}. For ease of reference, we refer to the aforementioned problem as \textit{GHZ-VM problem}. Although the GHZ-VM problem~\cite{JonHahTch-24,BriRau-01,FraLiuRai-23,ManPat-23} concerns the extraction of a single GHZ state, it offers a significant advantage over the Bell-VM formulation, since it inherently supports adaptability to the traffic requests. In fact, as mentioned above, a GHZ state allows the dynamic extraction of an EPR pair between any pair of nodes sharing the original GHZ state. And, notably, this extraction can occur at runtime, driven by the actual and potentially time-varying communication needs of the nodes.

It should be emphasized that the existing Bell-VM and GHZ-VM studies lie within the so-called vanilla extractions, as represented in Fig.~\ref{fig:01}. Specifically, there exists a subtle but key difference between remote and vanilla Gability/Pairability. In fact, in the latter, the extraction is performed regardless of the nodes proximity within the artificial topology. This in turn simplifies the problem with respect to constrain the extraction among nodes that are remote in the artificial topology. 

Accordingly, in this paper, we extend the Bell-VM and GHZ-VM approaches to a deeper problem, referred to as \textit{Remote-VM problem}, by determining the number of $n$-qubit GHZ states and Bell pairs, as well as the mass of GHZ that can be currently extracted among remote nodes of a given graph state, by using only single-qubit Clifford operations, single-qubit Pauli measurements, and classical communication.

\begin{figure}[t]
    \centering    \includegraphics[width=0.475\textwidth]{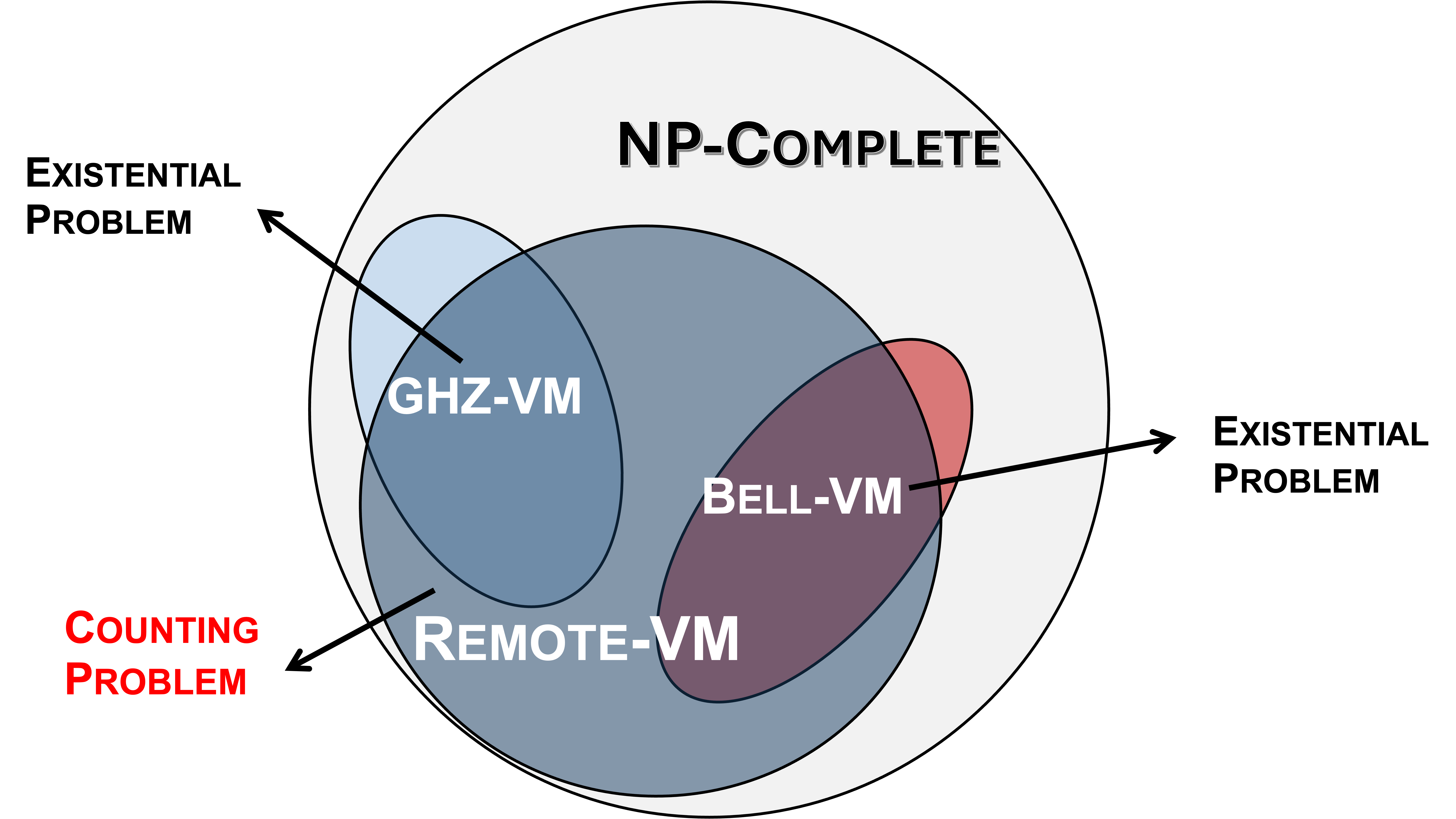}
    \caption{Venn diagram for the relationship between GHZ-VM, BELL-VM and REMOTE-VM (our research problem). }
    \label{fig:02}
    \hrulefill
\end{figure}

The difference between Remote-VM, existing Bell-VM and GHZ-VM is pictorially represented in Fig.~\ref{fig:02}. Specifically, while both Bell-VM and GHZ-VM are \textit{existential} decision problems (determining whether a resource state can be extracted), the Remote-VM problem belong to the \textit{counting problem class}. Thus, it not only resolve the existence, but also determine the exact extractable resources. It is important to note that Bell-VM and GHZ-VM have been proved to be NP-complete \cite{DahHelWeh-18,DahHelWeh-20}. Given that our problem introduces an additional constraint on top of Bell-VM and GHZ-VM, the Remote-VM problem is evidently at least in the NP-complete complexity class.

Specifically, the concurrent extraction of remote GHZ states and Bell pairs is equivalent to identifying disjoint independent sets in the original artificial topology. Leveraging this equivalence, the Remote-VM problem is reminiscent of the well-studied problem in the classical domain referred as $\# \text{IS}$ problem~\cite{DyeGre-00}, which is $\#$P-complete even when restricted to bipartite graphs~\cite{XiaZhaZha-07,CanPer-19}.
However, solving a counting problem in the quantum domain is not only linked to the structure of the graph, as in the classical $\# \text{IS}$ problem. Indeed the Remote-VM problem is constrained also by the operational limitations inherent to quantum systems. This additional constraint significantly increases the complexity of the problem, although it is out of the scope of this paper to study its complexity class.

With the above in mind, in the paper, in addition to the theoretical contribution highlighted in the Introduction, we also propose an heuristic solution for the NP-complete \textit{Remote-VM problem}. Specifically:
\begin{itemize} 
    \item We propose a polynomial-time algorithm for the Remote-VM (see Sec.~\ref{sec:4.1}), by exploiting graph-theory tools and only LC + LPM + CC. 
    \item The algorithm facilitates the extraction of both $n$-qubit GHZ states and EPR pairs among remote nodes, by extending the extraction capabilities beyond a set of apriori selected EPR pairs or a specific GHZ state. 
    \item The algorithm is able to provide the  three aforementioned parameters, namely volume, locations, and maximum mass, that rule the remote extraction.
    \item We evaluate the extractable volume for both remote Gability and remote Pairability, as well as the maximum mass for remote Gability, in general graph states. The analysis is conducted on representative Internet inspired artificial topologies and the results demonstrate their effectiveness (see Sec.~\ref{sec:5}).
\end{itemize}

To the best of our knowledge, this is the first paper rigorously investigating the extraction of remote entangled resources, while also providing an efficient polynomial-time procedure for its realization.

\section{Research Problem}
\label{sec:2}



We consider the worst case scenario, where  each qubit of the graph state is distributed to each network node. And we equivalently refer to node $i$ or to vertex $v_i$ associated with the qubit of the graph state $\ket{G}$ stored at such a node. To formally define our research problem, the following definitions are preliminary.

\begin{defin}[\textbf{Remote Nodes}]
    \label{def:01}
    Given a $N$-qubit graph state $\ket{G}$ and its corresponding graph $G=(V,E)$, two network nodes $i$ and $j$ are defined as \textit{remote} if the corresponding vertices $v_i,v_j$ are non-adjacent in $G$, i.e., if\footnote{In the following, with a small notation abuse, we denote un-directed edges as $(v_i,v_j)$ -- rather than with angle brackets as $\{v_i,v_j\}$ -- for notation simplicity.}:
    \begin{equation}
        \label{eq:01}
        (v_i,v_j) \not\in E.
    \end{equation}
\end{defin}

Throughout this paper, the notions of ``\textit{remoteness}'' and its counterpart, \textit{adjacency}, do not refer to the physical proximity of network nodes. Instead, they describe ``\textit{entangled proximity}'', that is, proximity within the artificial topology associated with the graph $G$ \cite{CacIllCal-23,CheIllCac-24}. Indeed, the presence of an edge within two vertices in $G$ represents an Ising interaction between the corresponding qubits of the graph state \cite{HeiEisBri-04,VanDurBri-08}. As a direct consequence of a graph state definition, it is evident that the graph associated to the graph state is a connected graph, meaning that a vertex is at least connected with another vertex. Two nodes are considered remote if they are not adjacent in this artificial topology. A subset of remote nodes is referred to ``\textit{Remote Subnet}'', as formally defined in Def.~\ref{def:02}.

\begin{defin}[\textbf{Remote Subnet}]
    \label{def:02}
    Given a $N$-qubit graph state $\ket{G}$ and its corresponding graph $G=(V,E)$, 
    a subset of network nodes $\tilde{V} \subset V$ is defined as \textit{remote subnet} if the following condition holds: 
    \begin{equation}
        \label{eq:02}
        \forall \, v_i,v_j \in \tilde{V} : (v_i,v_j) \not\in E.
    \end{equation}
\end{defin}

Stemming from the previous two definitions, we can now define the two main connectivity metrics. These two metrics focus on the quantum communication resources, i.e., GHZs and EPRs, that can be concurrently extracted among remote nodes in the artificial topology.

\begin{defin}[\textbf{$\bm{{r_g}(n)}$: remote $\bf{n}$-Gability}]
    \label{def:x03}
    The remote $n$-Gability of an $N$-qubit graph state $\ket{G}$ denotes the volume  of $n$-qubit GHZ states, with $n \leq N$, that can be eventually extracted among remote subnets via LC + LPM + CC operation. In the following, we denote the volume of $n$-qubit GHZ states as $r_g(n)$.
\end{defin} 

From Def.~\ref{def:x03}, it results that when it comes to GHZ states, there exist two dimensions that we must account for: the \textbf{volume}, similarly to EPRs, and the \textbf{mass} of each extracted GHZ, namely, the size of the GHZ in terms of qubit number. These two dimensions map into \textit{the number of the artificial subnets} that can be simultaneously extracted from the initial graph $\ket{G}$, and into \textit{the number of interconnected nodes in each subnet}. 
The third dimension, namely \textit{location}, is essential for both remote Pairability and $n$-Gability. Specifically, it plays a critical role in leveraging entangled resources by enabling the unambiguous identification of the participating nodes within the quantum states.

\begin{remark}
    Since an EPR pair is a special case of a GHZ state with two qubits, the case of $r_g(2)$ is essentially a special instance of remote $n$-Gability. We refer to this as remote Pairability, which is formally defined in Def.~\ref{def:04}.
\end{remark}

\begin{defin}[\textbf{$\bm{r_e}$: remote Pairability}]
    \label{def:04}
    The remote Pairability of an  $N$-qubit graph state $\ket{G}$ denotes the volume of EPR pairs that can be eventually extracted between remote nodes via LC + LPM + CC operation. 
    In the following, we denote the volume of EPRs as $r_e$, by omitting the dependence on $\ket{G}$ for the sake of notation brevity.
\end{defin} 

It is worthwhile to mention that solving the Gability problem is more difficult than solving the Pairability problem, and the following inequality holds for the volume whenever $n > 2$:
\begin{equation}
    \label{eq:03}
    r_g(n) \leq r_g(n-1) \leq r_g(2) \eqdef r_e
\end{equation}

Stemming from the concept of remote  $n$-Gability and Pairability in Defs.~\ref{def:x03} and~\ref{def:04}, we can now formally define our research problem.


\begin{figure}[t]
    \centering    \includegraphics[width=0.5\textwidth]{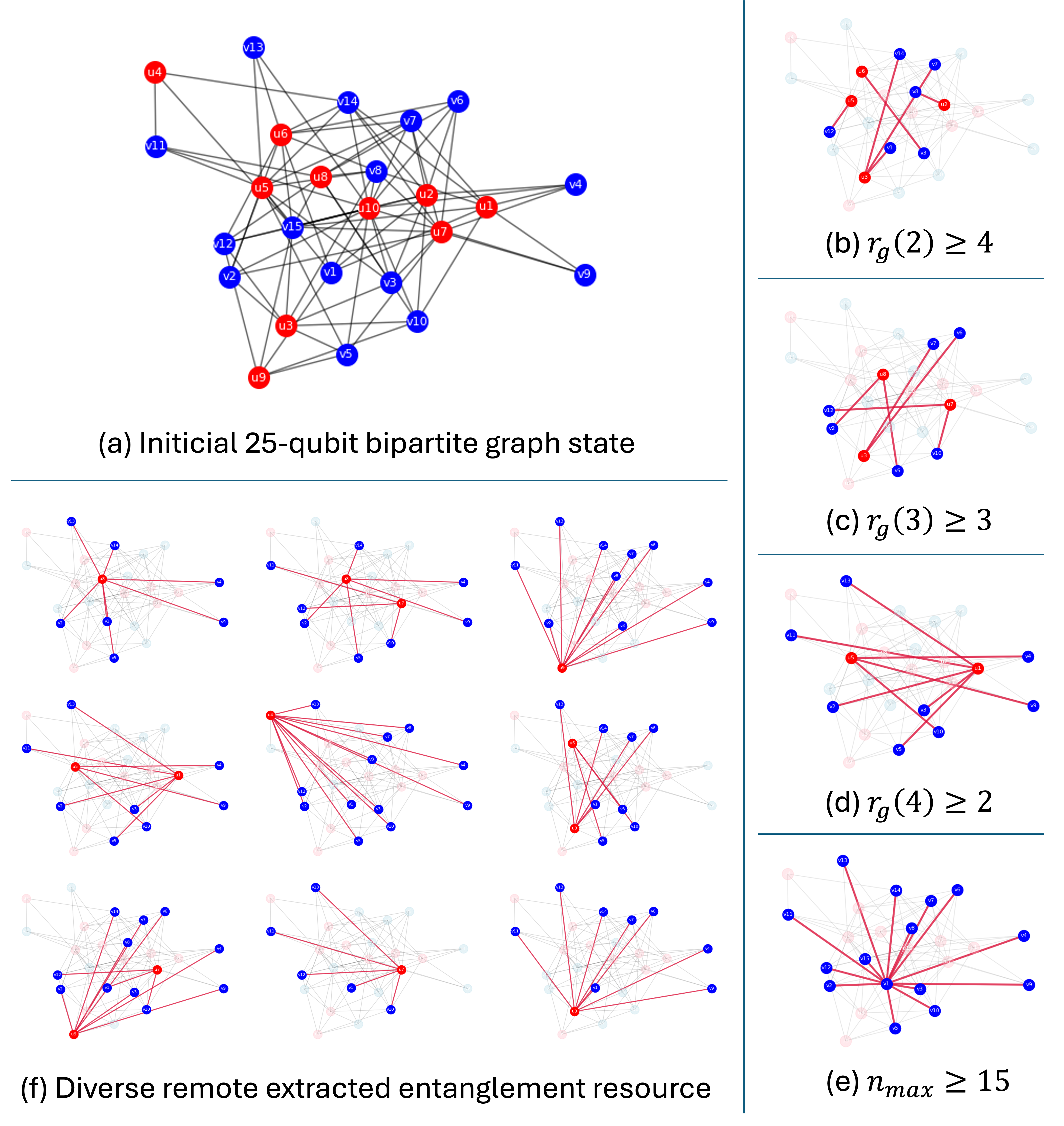}
    \caption{Pictorial representation of the research problem. (a) The initial 25-qubit bipartite graph state $\ket{G}$. The goal is to constructively address the Remote-VM problem: determining the bound for the volume of extractions can be performed simultaneously from $\ket{G}$, and the bound for the maximum mass of remote GHZ states that can be extracted, and identifying (location) the remote nodes involved in these extractions. (b) A solution to the Remote Pairability problem, identifying pairs of remote nodes that can be entangled. (c–d) Examples for the Remote $n$-Gability problem, with $n=3,4$. (e) Extraction of a 15-qubit remote GHZ state, representing the lower bound of maximum achievable mass of a remote GHZ state from the initial graph. (f) Illustration of diverse extracted remote resources obtained from $\ket{G}$.}
    \label{fig:04}
    \hrulefill
\end{figure}

\begin{prob} [Remote-VM problem]
    Given a graph state $\ket{G}$ distributed across the network nodes, we determine the  volume, location and maximum mass of $n$-qubit GHZ states and EPR pairs, extractable among remote nodes, by using only single-qubit Clifford operations, single-qubit Pauli measurements and classical communications.
\end{prob}

As aforementioned, this problem is NP-complete. Thus, we theoretically derive bounds for remote pairing and $n$-ability for an arbitrary graph state. Specifically, our goal is to determine bounds for:
    \begin{itemize}
        \item[i)] the volume $r_g(n)$  for any value of $n$, as well as the locations  of the remote nodes eventually sharing the GHZ states;
        \item[ii)] the mass of the remote $n$-Gability; 
        \item[iii)] and the volume $r_e$ of the remote Pairability, as well as the locations  of the remote nodes eventually sharing the EPR pairs.       
    \end{itemize}

We underline that the derived lower bounds are far from being only theoretical, since they are derived by individuating the locations of the nodes that share the extracted EPRs/GHZs. Hence, these bounds are \textit{constructive} in the sense that not only they determine whether a solution exists, but they also construct the solution explicitly.

In Fig.~\ref{fig:04}, we provide a pictorial representation of the formulated research problem to better grasp the implications of the remote extractions from a network perspective.


\section{Remote Pairability and Remote $n$-Gability}
\label{sec:3}

Here, we first provide some preliminaries in Sec.~\ref{sec:3.1}. Then, in Sec.~\ref{sec:3.2} we derive the remote extraction conditions for both remote $n$-Gability and \textit{Pairability} for two-colorable graph state $\ket{G}$, in Lemmas~\ref{lem:01} and \ref{lem:02}, respectively. And we also provide the bound for maximum mass $n_{max}$ for remote G-ability in Lemma~\ref{lem:03}.

\subsection{Preliminaries}
\label{sec:3.1}

We first introduce several fundamental definitions from graph theory, including the concepts of maximum degree and maximum independent set. These metrics play a crucial role in characterizing entangled proximity in artificial topologies and directly influence the achievable $n$-Gability and Pairability in the Remote-VM problem.

\begin{defin}[\textbf{Maximum degree}]
    \label{def:05}
    The maximum degree of a graph $G=(V,E)$, denoted by $\Delta(G)$, is the largest degree of any vertex in $V$:
    \begin{equation}
        \label{eq:04}
        \Delta(G) = \max\{ \deg(v) \mid v \in V \}
    \end{equation}
    with $\deg(v)$ is the number of neighbors of vertex $v$ in $G$.
\end{defin}

\begin{defin}[\textbf{Maximum Independent set}]
    \label{def:06}
    A maximum independent set is the independent set of largest  size for a given graph $G$. This size is called the independence number of $G$, denoted by $\alpha(G)$.
\end{defin}

\begin{remark}
    Remote $n$-Gability relies on identifying independent sets of size $n$ in the artificial topology, as each of such sets may enable the extraction of a remote $n$-qubit GHZ state. The larger is the independent set permitted by the artificial topology, the greater is the mass of the extracted  remote GHZ state. As a result, the size $\alpha(G)$ of the maximum independent set determines the theoretical upper bound for the mass of a remote GHZ states. 
\end{remark}

As aforementioned, we focus on two-colorable\footnote{In principle, coloring assigns colors to arbitrary elements of a graph according to arbitrary partition constrains. In the following, we adopt the most widely-used partition constraint based on vertex adjacency, since other coloring problems can be easily transformed into a vertex coloring problem.} graph states. This choice is not restrictive, since any graph state can be converted in a two-colorable one under relaxed conditions\footnote{We acknowledge that converting arbitrary graph states to bipartite form generally entails overhead, as discussed in Section~\ref{sec:6}.}\cite{VanDehDe-04,HeiDurEis-06}. Furthermore, two-colorable graphs model a wide range of important communication network topologies, such as butterfly, bistar, tree, linear, even loop, grid, star, cluster networks, highly exploited in entanglement-based communication protocols\cite{TanZhaKra-23,CheIllCac-24,FreDur-24,HahPapEis-19,BraShaSze-22,MazCalCac-24,BenHajVan-24,AzuTamLo-15,MazZhaChu-25}. In addition, two-colorable graph states are local-unitary (LU) equivalent to Calderbank-Shor-Steane (CSS) states, which are important in quantum error correction strategies \cite{cheLo-08,CalSho-96,Ste-96}.  Formally, we have the following definition.

\begin{defin}[\textbf{Two-colorable Graph} or \textbf{Bipartite Graph}]
    \label{def:07}
    A graph $G=(V,E)$ is two-colorable if the set of vertices $V$ can be partitioned
    into two subsets $\{ P_1, P_2 \}$ so that there exist no edge in $E$ between two vertices belonging to the same subset. Two-colorable graph $G=(V,E)$ can be also denoted as $G=(P_1, P_2, E)$.
\end{defin}


\begin{defin}[\textbf{Star vertex}]
    \label{def:08}
    Given a two-colorable graph $G = (P_1, P_2,E)$, the vertex $v_i$ belonging to partition $P_i$ is defined as star vertex if its neighborhood $N(v_i)$ coincides with the opposite partition $P_j\eqdef V \setminus P_i$, i.e.,:
    \begin{align}
        \label{eq:05}
         N(v_i) \eqdef \big\{ v_j \in V : (v_i,v_j) \in E \big \} \equiv V \setminus P_i \eqdef P_j.
    \end{align}
\end{defin}

\begin{remark}
    We underline that our definition of star vertex is not the common one used in graph theory, where a star vertex denotes a vertex connected to all the other vertices in $V$. In fact, our definition is related to the vertex partitioning, and thus, our star vertex undergoes the coloring constraint. Consequently, the star vertex is not connected to the vertices belonging to its own partition. 
\end{remark}

In the following, for the sake of notation simplicity, we denoted with $S_1 \subseteq P_1$ and $S_2 \subseteq P_2$ the set of star vertices in the two partitions, i.e.:
\begin{align}
    \label{eq:06}
    S_1 &= \big\{ v_i \in P_1 : N(v_i) = P_2 \big\}, \\
    \label{eq:07}
    S_2 &= \big\{ v_j \in P_2 : N(v_j) = P_1 \big\},
\end{align}
and, accordingly, by denoting the remaining vertices, i.e. non-star vertices, in each partition as $V_1$ and $V_2$, we can adopt the following labeling for the two-colorable graph $G=(P_1,P_2,E)$:
\begin{align}
    \label{eq:08}
    P_1 = S_1 \cup V_1 \\
    \quad \text{ with } S_1 &= \{s^1_1,\cdots, s^{k_1}_1\} \wedge  V_1 = \{v^1_1,\cdots,v^{n_1}_1\}, \nonumber \\
    \label{eq:09}
    P_2 = S_2 \cup V_2 \\
    \quad \text{ with } S_2 &= \{s^1_2,\cdots, s_2^{k_2} \} \wedge  V_2 = \{v^1_2,\cdots,v^{n_2}_2\}, \nonumber
 \end{align}
with $|P_1| = n_1 + k_1$ and $|P_2| = n_2 + k_2 $.

\begin{defin}[\textbf{Opposite Remote-Set}]
    \label{def:09}
    Given a two-colorable graph $G=(P_1, P_2, E)$, the \textit{opposite remote set} of the arbitrary vertex $v_i \in P_i$, with $i \in \{1,2\}$, is the set $\overline{N}(v_i)$ of remote vertices of $v_i$ belonging to the other partition:
    \begin{equation}
        \label{eq:10}
        \overline{N}(v_i) \eqdef \big\{ v_j \in P_j \neq P_i : (v_i,v_j ) \not\in E \big\}.
    \end{equation}
\end{defin}
The term ``opposite'' in Def.~\ref{def:09} is used to highlight that the remote nodes belong to different partitions. This will be exploited in the next sections for carrying the theoretical analysis. Clearly, vertices belonging to the same partition are remote \textit{per se}, as a consequence of the two-colorable graph state definition. The concept of the opposite remote set can be extended from individual vertices to subsets of vertices within the same partition. For any subset $A \subseteq P_i$, the union and intersection opposite remote sets are defined respectively as:
\begin{align}
    \label{eq:11}
    \overline{N}_{\cup}(A) &\eqdef \bigcup_{v_i \in A} \overline{N}(v_i) \\
    \label{eq:12}
    \overline{N}_{\cap}(A) &\eqdef \bigcap_{v_i \in A} \overline{N}(v_i)
\end{align}

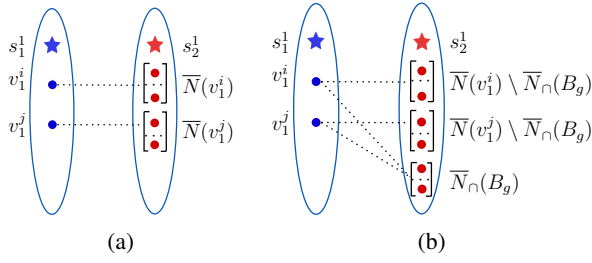
\begin{figure}[t]
    \centering
    \begin{minipage}[b]{0.5\textwidth}
    \centering
        \begin{minipage}[b]{0.37\textwidth}
            \centering\resizebox{\textwidth}{!}{
        \tikzset{every picture/.style={line width=0.75pt}} 

\begin{tikzpicture}[x=0.75pt,y=0.75pt,yscale=-1,xscale=1]

\draw  [color={rgb, 255:red, 12; green, 90; blue, 182 }  ,draw opacity=1 ] (63,110.28) .. controls (63,66.15) and (70.66,30.38) .. (80.1,30.38) .. controls (89.54,30.38) and (97.2,66.15) .. (97.2,110.28) .. controls (97.2,154.41) and (89.54,190.18) .. (80.1,190.18) .. controls (70.66,190.18) and (63,154.41) .. (63,110.28) -- cycle ;
\draw  [color={rgb, 255:red, 255; green, 255; blue, 255 }  ,draw opacity=0.38 ][fill={rgb, 255:red, 54; green, 58; blue, 230 }  ,fill opacity=1 ][line width=0.75]  (80.3,50.18) -- (82.68,55.42) -- (88,56.27) -- (84.15,60.35) -- (85.06,66.12) -- (80.3,63.39) -- (75.54,66.12) -- (76.45,60.35) -- (72.6,56.27) -- (77.92,55.42) -- cycle ;
\draw  [color={rgb, 255:red, 255; green, 255; blue, 255 }  ,draw opacity=0.44 ][fill={rgb, 255:red, 12; green, 2; blue, 208 }  ,fill opacity=1 ] (76.6,89.5) .. controls (76.6,87.35) and (78.35,85.6) .. (80.5,85.6) .. controls (82.65,85.6) and (84.4,87.35) .. (84.4,89.5) .. controls (84.4,91.65) and (82.65,93.4) .. (80.5,93.4) .. controls (78.35,93.4) and (76.6,91.65) .. (76.6,89.5) -- cycle ;
\draw  [color={rgb, 255:red, 255; green, 255; blue, 255 }  ,draw opacity=0.44 ][fill={rgb, 255:red, 12; green, 2; blue, 208 }  ,fill opacity=1 ] (76.6,120.5) .. controls (76.6,118.35) and (78.35,116.6) .. (80.5,116.6) .. controls (82.65,116.6) and (84.4,118.35) .. (84.4,120.5) .. controls (84.4,122.65) and (82.65,124.4) .. (80.5,124.4) .. controls (78.35,124.4) and (76.6,122.65) .. (76.6,120.5) -- cycle ;
\draw  [color={rgb, 255:red, 12; green, 90; blue, 182 }  ,draw opacity=1 ] (143,109.94) .. controls (143,65.82) and (150.66,30.04) .. (160.1,30.04) .. controls (169.54,30.04) and (177.2,65.82) .. (177.2,109.94) .. controls (177.2,154.07) and (169.54,189.84) .. (160.1,189.84) .. controls (150.66,189.84) and (143,154.07) .. (143,109.94) -- cycle ;
\draw  [color={rgb, 255:red, 255; green, 255; blue, 255 }  ,draw opacity=0.38 ][fill={rgb, 255:red, 230; green, 54; blue, 54 }  ,fill opacity=1 ][line width=0.75]  (160.3,49.84) -- (162.68,55.09) -- (168,55.93) -- (164.15,60.02) -- (165.06,65.78) -- (160.3,63.06) -- (155.54,65.78) -- (156.45,60.02) -- (152.6,55.93) -- (157.92,55.09) -- cycle ;
\draw  [color={rgb, 255:red, 255; green, 255; blue, 255 }  ,draw opacity=0.44 ][fill={rgb, 255:red, 208; green, 2; blue, 2 }  ,fill opacity=1 ] (156.35,80.42) .. controls (156.35,78.26) and (158.1,76.52) .. (160.25,76.52) .. controls (162.4,76.52) and (164.15,78.26) .. (164.15,80.42) .. controls (164.15,82.57) and (162.4,84.32) .. (160.25,84.32) .. controls (158.1,84.32) and (156.35,82.57) .. (156.35,80.42) -- cycle ;
\draw  [color={rgb, 255:red, 255; green, 255; blue, 255 }  ,draw opacity=0.44 ][fill={rgb, 255:red, 208; green, 2; blue, 2 }  ,fill opacity=1 ] (156.35,99.67) .. controls (156.35,97.51) and (158.1,95.77) .. (160.25,95.77) .. controls (162.4,95.77) and (164.15,97.51) .. (164.15,99.67) .. controls (164.15,101.82) and (162.4,103.57) .. (160.25,103.57) .. controls (158.1,103.57) and (156.35,101.82) .. (156.35,99.67) -- cycle ;
\draw  [color={rgb, 255:red, 255; green, 255; blue, 255 }  ,draw opacity=0.44 ][fill={rgb, 255:red, 208; green, 2; blue, 2 }  ,fill opacity=1 ] (156.1,119.17) .. controls (156.1,117.01) and (157.85,115.27) .. (160,115.27) .. controls (162.15,115.27) and (163.9,117.01) .. (163.9,119.17) .. controls (163.9,121.32) and (162.15,123.07) .. (160,123.07) .. controls (157.85,123.07) and (156.1,121.32) .. (156.1,119.17) -- cycle ;
\draw  [color={rgb, 255:red, 255; green, 255; blue, 255 }  ,draw opacity=0.44 ][fill={rgb, 255:red, 208; green, 2; blue, 2 }  ,fill opacity=1 ] (156.1,136.17) .. controls (156.1,134.01) and (157.85,132.27) .. (160,132.27) .. controls (162.15,132.27) and (163.9,134.01) .. (163.9,136.17) .. controls (163.9,138.32) and (162.15,140.07) .. (160,140.07) .. controls (157.85,140.07) and (156.1,138.32) .. (156.1,136.17) -- cycle ;
\draw   (165.75,73.19) -- (169,73.19) -- (169,103.69) ;
\draw    (169,103.69) -- (164.75,103.69) ;

\draw   (154.98,103.44) -- (151.73,103.43) -- (151.77,72.93) ;
\draw    (151.77,72.93) -- (156.02,72.94) ;

\draw   (165.75,110.69) -- (169,110.69) -- (169,141.19) ;
\draw    (169,141.19) -- (164.75,141.19) ;

\draw   (154.98,140.94) -- (151.73,140.93) -- (151.77,110.43) ;
\draw    (151.77,110.43) -- (156.02,110.44) ;

\draw  [dash pattern={on 0.84pt off 2.51pt}]  (84.4,89.5) -- (151,89.94) ;
\draw  [dash pattern={on 0.84pt off 2.51pt}]  (84.4,120.5) -- (151.5,120.69) ;

\draw (150,85) node [anchor=north west][inner sep=0.75pt]    {$\cdots $};
\draw (150,125) node [anchor=north west][inner sep=0.75pt]    {$\cdots $};

\draw (45,50) node [anchor=north west][inner sep=0.75pt]  [font=\large] [align=left] {$s^1_1$};

\draw (45,75) node [anchor=north west][inner sep=0.75pt]  [font=\large] [align=left] {$v^i_1$};

\draw (45,110) node [anchor=north west][inner sep=0.75pt]  [font=\large] [align=left] {$v^j_1$};

\draw (180,50) node [anchor=north west][inner sep=0.75pt]  [font=\large] [align=left] {$s^1_2$};

\draw (180,80) node [anchor=north west][inner sep=0.75pt]  [font=\large] [align=left] {$\overline{N}(v^i_1)$};

\draw (180,115) node [anchor=north west][inner sep=0.75pt]  [font=\large] [align=left] {$\overline{N}(v^j_1)$};


\end{tikzpicture}
    }
            \subcaption{}
    	   \label{fig:05.1}
        \end{minipage}
        \begin{minipage}[b]{0.51\textwidth}
            \centering\resizebox{\textwidth}{!}{
        \tikzset{every picture/.style={line width=0.75pt}} 

\begin{tikzpicture}[x=0.75pt,y=0.75pt,yscale=-1,xscale=1]

\draw  [color={rgb, 255:red, 12; green, 90; blue, 182 }  ,draw opacity=1 ] (263,110.28) .. controls (263,66.15) and (270.66,30.38) .. (280.1,30.38) .. controls (289.54,30.38) and (297.2,66.15) .. (297.2,110.28) .. controls (297.2,154.41) and (289.54,190.18) .. (280.1,190.18) .. controls (270.66,190.18) and (263,154.41) .. (263,110.28) -- cycle ;
\draw  [color={rgb, 255:red, 12; green, 90; blue, 182 }  ,draw opacity=1 ] (343,109.94) .. controls (343,65.82) and (350.66,30.04) .. (360.1,30.04) .. controls (369.54,30.04) and (377.2,65.82) .. (377.2,109.94) .. controls (377.2,154.07) and (369.54,189.84) .. (360.1,189.84) .. controls (350.66,189.84) and (343,154.07) .. (343,109.94) -- cycle ;
\draw  [color={rgb, 255:red, 255; green, 255; blue, 255 }  ,draw opacity=0.38 ][fill={rgb, 255:red, 54; green, 58; blue, 230 }  ,fill opacity=1 ][line width=0.75]  (279.85,50.18) -- (282.23,55.42) -- (287.55,56.27) -- (283.7,60.35) -- (284.61,66.12) -- (279.85,63.39) -- (275.09,66.12) -- (276,60.35) -- (272.15,56.27) -- (277.47,55.42) -- cycle ;
\draw  [color={rgb, 255:red, 255; green, 255; blue, 255 }  ,draw opacity=0.44 ][fill={rgb, 255:red, 12; green, 2; blue, 208 }  ,fill opacity=1 ] (276.15,89.5) .. controls (276.15,87.35) and (277.9,85.6) .. (280.05,85.6) .. controls (282.2,85.6) and (283.95,87.35) .. (283.95,89.5) .. controls (283.95,91.65) and (282.2,93.4) .. (280.05,93.4) .. controls (277.9,93.4) and (276.15,91.65) .. (276.15,89.5) -- cycle ;
\draw  [color={rgb, 255:red, 255; green, 255; blue, 255 }  ,draw opacity=0.44 ][fill={rgb, 255:red, 12; green, 2; blue, 208 }  ,fill opacity=1 ] (276.15,120.5) .. controls (276.15,118.35) and (277.9,116.6) .. (280.05,116.6) .. controls (282.2,116.6) and (283.95,118.35) .. (283.95,120.5) .. controls (283.95,122.65) and (282.2,124.4) .. (280.05,124.4) .. controls (277.9,124.4) and (276.15,122.65) .. (276.15,120.5) -- cycle ;
\draw  [dash pattern={on 0.84pt off 2.51pt}]  (283.95,89.5) -- (350.55,89.94) ;
\draw  [dash pattern={on 0.84pt off 2.51pt}]  (283.95,120.5) -- (351.05,120.69) ;
\draw  [color={rgb, 255:red, 255; green, 255; blue, 255 }  ,draw opacity=0.38 ][fill={rgb, 255:red, 230; green, 54; blue, 54 }  ,fill opacity=1 ][line width=0.75]  (360.3,50.84) -- (362.68,56.09) -- (368,56.93) -- (364.15,61.02) -- (365.06,66.78) -- (360.3,64.06) -- (355.54,66.78) -- (356.45,61.02) -- (352.6,56.93) -- (357.92,56.09) -- cycle ;
\draw  [color={rgb, 255:red, 255; green, 255; blue, 255 }  ,draw opacity=0.44 ][fill={rgb, 255:red, 208; green, 2; blue, 2 }  ,fill opacity=1 ] (356.35,81.42) .. controls (356.35,79.26) and (358.1,77.52) .. (360.25,77.52) .. controls (362.4,77.52) and (364.15,79.26) .. (364.15,81.42) .. controls (364.15,83.57) and (362.4,85.32) .. (360.25,85.32) .. controls (358.1,85.32) and (356.35,83.57) .. (356.35,81.42) -- cycle ;
\draw  [color={rgb, 255:red, 255; green, 255; blue, 255 }  ,draw opacity=0.44 ][fill={rgb, 255:red, 208; green, 2; blue, 2 }  ,fill opacity=1 ] (356.35,100.67) .. controls (356.35,98.51) and (358.1,96.77) .. (360.25,96.77) .. controls (362.4,96.77) and (364.15,98.51) .. (364.15,100.67) .. controls (364.15,102.82) and (362.4,104.57) .. (360.25,104.57) .. controls (358.1,104.57) and (356.35,102.82) .. (356.35,100.67) -- cycle ;
\draw  [color={rgb, 255:red, 255; green, 255; blue, 255 }  ,draw opacity=0.44 ][fill={rgb, 255:red, 208; green, 2; blue, 2 }  ,fill opacity=1 ] (356.2,155.54) .. controls (356.2,153.39) and (357.95,151.64) .. (360.1,151.64) .. controls (362.25,151.64) and (364,153.39) .. (364,155.54) .. controls (364,157.7) and (362.25,159.44) .. (360.1,159.44) .. controls (357.95,159.44) and (356.2,157.7) .. (356.2,155.54) -- cycle ;
\draw  [color={rgb, 255:red, 255; green, 255; blue, 255 }  ,draw opacity=0.44 ][fill={rgb, 255:red, 208; green, 2; blue, 2 }  ,fill opacity=1 ] (356.53,171.04) .. controls (356.53,168.89) and (358.28,167.14) .. (360.43,167.14) .. controls (362.59,167.14) and (364.33,168.89) .. (364.33,171.04) .. controls (364.33,173.2) and (362.59,174.94) .. (360.43,174.94) .. controls (358.28,174.94) and (356.53,173.2) .. (356.53,171.04) -- cycle ;
\draw  [color={rgb, 255:red, 255; green, 255; blue, 255 }  ,draw opacity=0.44 ][fill={rgb, 255:red, 208; green, 2; blue, 2 }  ,fill opacity=1 ] (356.1,120.17) .. controls (356.1,118.01) and (357.85,116.27) .. (360,116.27) .. controls (362.15,116.27) and (363.9,118.01) .. (363.9,120.17) .. controls (363.9,122.32) and (362.15,124.07) .. (360,124.07) .. controls (357.85,124.07) and (356.1,122.32) .. (356.1,120.17) -- cycle ;
\draw  [color={rgb, 255:red, 255; green, 255; blue, 255 }  ,draw opacity=0.44 ][fill={rgb, 255:red, 208; green, 2; blue, 2 }  ,fill opacity=1 ] (356.1,137.17) .. controls (356.1,135.01) and (357.85,133.27) .. (360,133.27) .. controls (362.15,133.27) and (363.9,135.01) .. (363.9,137.17) .. controls (363.9,139.32) and (362.15,141.07) .. (360,141.07) .. controls (357.85,141.07) and (356.1,139.32) .. (356.1,137.17) -- cycle ;
\draw   (365.75,74.19) -- (369,74.19) -- (369,104.69) ;
\draw    (369,104.69) -- (364.75,104.69) ;

\draw   (354.98,104.44) -- (351.73,104.43) -- (351.77,73.93) ;
\draw    (351.77,73.93) -- (356.02,73.94) ;

\draw   (365.75,111.69) -- (369,111.69) -- (369,142.19) ;
\draw    (369,142.19) -- (364.75,142.19) ;

\draw   (354.98,141.94) -- (351.73,141.93) -- (351.77,111.43) ;
\draw    (351.77,111.43) -- (356.02,111.44) ;

\draw   (364.25,151.19) -- (367.5,151.19) -- (367.5,176.44) ;
\draw    (367.5,176.44) -- (363.25,176.44) ;

\draw   (355.48,176.19) -- (352.23,176.18) -- (352.26,150.94) ;
\draw    (352.26,150.94) -- (356.51,150.94) ;

\draw  [dash pattern={on 0.84pt off 2.51pt}]  (283.95,89.5) -- (351.05,161.69) ;
\draw  [dash pattern={on 0.84pt off 2.51pt}]  (283.95,120.5) -- (351.05,161.69) ;

\draw (350,85) node [anchor=north west][inner sep=0.75pt]    {$\cdots $};
\draw (350,125) node [anchor=north west][inner sep=0.75pt]    {$\cdots $};
\draw (350,160) node [anchor=north west][inner sep=0.75pt]    {$\cdots $};

\draw (245,50) node [anchor=north west][inner sep=0.75pt]  [font=\large] [align=left] {$s^1_1$};

\draw (245,75) node [anchor=north west][inner sep=0.75pt]  [font=\large] [align=left] {$v^i_1$};

\draw (245,110) node [anchor=north west][inner sep=0.75pt]  [font=\large] [align=left] {$v^j_1$};

\draw (380,50) node [anchor=north west][inner sep=0.75pt]  [font=\large] [align=left] {$s^1_2$};

\draw (380,80) node [anchor=north west][inner sep=0.75pt]  [font=\large] [align=left] {$\overline{N}(v^i_1)\setminus \overline{N}_{\cap}(B_g)$};

\draw (380,115) node [anchor=north west][inner sep=0.75pt]  [font=\large] [align=left] {$\overline{N}(v^j_1)\setminus \overline{N}_{\cap}(B_g)$};

\draw (380,155) node [anchor=north west][inner sep=0.75pt]  [font=\large] [align=left] {$\overline{N}_{\cap}(B_g)$};


\end{tikzpicture}
    }
            \subcaption{}
    	   \label{fig:05.2}
        \end{minipage}
    \end{minipage}
    \caption{Pictorial representation for the conditions of Lemmas~\ref{lem:01} and~\ref{lem:02}. Dashed lines denote opposite remote sets (Def.\ref{def:09}), where a dashed line from a vertex to a set of vertices in squared parentheses indicates remote subnets. (a) In $G$, vertices $v^i_1$ and $v^j_1$ in $V_1$ satisfy \eqref{eq:15}, so $A_g = \{v^i_1, v^j_1\}$ by Lem.\ref{lem:01}. (b) In $G$, $v^i_1$ and $v^j_1$ in $V_1$ satisfy \eqref{eq:16}, so $B_g = \{v^i_1, v^j_1\}$ by Lem.~\ref{lem:02}.} 
    \label{fig:05}
    \hrulefill
\end{figure}

\subsection{Remote Extraction Conditions}
\label{sec:3.2}
We provide two  sufficient conditions in Lem.~\ref{lem:01} and Lem.~\ref{lem:02} for remote $n$-Gability in two-colorable graph states 
$\ket{G}$, where remote Pairability is treated as the special case of $n$-Gability for $n=2$. And,
a pictorial representation of the aforementioned conditions is given in Fig.~\ref{fig:05}.

\begin{lem}[\textbf{Remote $n$-Gability: Condition I}]
    \label{lem:01}
    Let $\ket{G}$ be a two-colorable graph state, with corresponding graph $G=(P_1, P_2, E)$. A sufficient condition for concurrently extracting $\dot{r}_g(n)$ GHZ states, each involving $n$ qubits, is that $\dot{r}_g(n)$ vertices in one partition have pairwise disjoint opposite remote sets of cardinality at least $n - 1$, and that there exists at least one star vertex in each partition. Formally: 
    \begin{align}
        \label{eq:13} 
        & \exists S_1, S_2 \neq \emptyset, \exists A_g \subseteq V_i, \text{with}\, |A_g|=\dot{r}_g(n): \\
        &\nonumber |\overline{N}(v_i) | \geq n - 1, \forall v_i \in A_g  \; \; \wedge \\
        &\nonumber \overline{N}(v_i) \cap \overline{N}(v_j) \equiv \emptyset, 
       \forall v_i, v_j \in A_g, v_i \neq v_j.  
    \end{align}
    \begin{IEEEproof} 
    Please refer to App.~\ref{app:lem:01}. 
    \end{IEEEproof}
\end{lem}

Since an EPR pair can be regarded as a degenerate case of a GHZ state involving two qubits, the sufficient condition in Lemma~\ref{lem:01} with $n = 2$ applies directly to the concurrent extraction of $\dot{r}_g(2)$ EPR pairs. This result is formally stated in Corollary~\ref{cor:01}.

\begin{cor}[\textbf{Remote Pairability: Condition I}]
    \label{cor:01}
    Let $\ket{G}$ be a two-colorable graph state, with corresponding graph $G=(P_1, P_2, E)$, and let $A_g$ denote the set defined in \eqref{eq:13}, for $n=2$. A sufficient condition for concurrently extracting $\dot{r}_g(2)$ EPR pairs among remote nodes is that  $\dot{r}_g(2)$ vertices in $A_g$ have disjoint opposite sets of cardinality at least 1, and that there exists at least one star vertex in each partition. 
\begin{IEEEproof} 
    The proof follows by reasoning as in App.~\ref{app:lem:01} for $n=2$. 
    \end{IEEEproof}\end{cor}

\begin{lem}[\textbf{Remote $n$-Gability: Condition II}]
    \label{lem:02}
    Let $\ket{G}$ be a two-colorable graph state, with corresponding graph $G=(P_1, P_2, E)$. A sufficient condition for concurrently extracting $\ddot{r}_g(n)$ $n$-GHZ states is that $\ddot{r}_g(n)$ vertices in one partition have opposite remote sets that share only one unique intersection, with each opposite remote set retaining at least $(n-1)$ vertices after excluding the common intersection, and that there exists at least one star vertex in each partition. Formally:
    \begin{align}
        \label{eq:14}
        & \exists S_1, S_2 \neq \emptyset, \exists! \overline{N}_{\cap}(B_g),\, \text{with} \, B_g \subseteq V_i \,\text{and}\, |B_g|=\ddot{r}_g(n) : \\&\nonumber
         |\overline{N}(v_i) \setminus \overline{N}_{\cap}(B_g)| \geq n-1, \forall v_i \in B_g \; \wedge \\
        & \nonumber \left( \overline{N}(v_i) \setminus \overline{N}_{\cap}(B_g) \right) \cap \left( \overline{N}(v_j) \setminus \overline{N}_{\cap}(B_g) \right) \equiv \emptyset, \forall v_i, v_j \in B_g, v_i \neq v_j.
    \end{align}
with $\overline{N}_{\cap}(B_g) \subset P_j \neq P_i$  defined in \eqref{eq:12}.
    \begin{IEEEproof}
        By removing $\overline{N}_{\cap}(B_g)$, the vertices in $B_g$ will satisfy Lem.~\ref{lem:01}, becoming as $A_g$. Please refer to App.~\ref{app:lem:01}.  
    \end{IEEEproof}
\end{lem}

Similarly, the sufficient condition in Lemma~\ref{lem:02} with $n = 2$ applies directly to the concurrent extraction of $\ddot{r}_g(2)$ EPR pairs. This result is formally stated in Corollary~\ref{cor:02}.

\begin{cor}[\textbf{Remote Pairability: Condition II}]
    \label{cor:02}
    Let $\ket{G}$ be a two-colorable graph state, with corresponding graph $G=(P_1, P_2, E)$ and let $B_g$ denote the set defined in \eqref{eq:14} for $n=2$. A sufficient condition for concurrently extracting $\ddot{r}_g(2)$ EPR pairs, is that there exist $\ddot{r}_g(2)$ vertices in $B_g$ and that there exists at least one star vertex in each partition. 
\end{cor}

For an arbitrary graph state, it may happen that only one partition or no partition in $\ket{G}$ contains star vertices. Thus, neither Lemmas~\ref{lem:01} nor \ref{lem:02} can be exploited to assess $n$-Gability and pairability. Here, we work toward such an issue by introducing additional graph manipulations, as formally defined in Cor.~\ref{cor:03}.

\begin{cor}
    \label{cor:03}
    Let $\ket{G}$ be a two-colorable graph state, with corresponding graph $G=(P_1, P_2, E)$, where $P_j=V_j$, namely $P_j$ contains no star vertices. $G$ can be reduced to a graph $G'$ -- vertex minor of $G$ -- characterized by a star vertex in partition $P_j$, as follows:
    \begin{equation}
        \label{eq:15}
        G' = G \setminus \overline{N}(v^i_j)
    \end{equation}
    with $v^i_j \in P_j$ denoting the new star vertex.
    \begin{IEEEproof}
        By removing the opposite remote-set of $v^i_j$, the neighborhood $N(v^i_j)$ in the resulting graph $G'$ coincides with $V_{\bar{j}}$. Hence $v^i_j$ becomes a star vertex in $P_j$ according to Def.~\ref{def:08}. 
    \end{IEEEproof}
\end{cor}

By exploiting Cor.~\ref{cor:03}, each partition of a general graph state can be forced to contain at least one star vertex. This structural property serves as a necessary condition for both Lemma~\ref{lem:01} and Lemma~\ref{lem:02}. With this in mind, let us denote with $\mathcal{A}_g$ the collections of sets satisfying \eqref{eq:13}, while  $\tilde{\mathcal{A}}_g \subseteq \mathcal{A}_g$ comprises the maximal-cardinality subsets of size $\tilde{\dot{r}}_g(n)$, i.e.:
\begin{equation}
    \label{eq:016}
    \tilde{\mathcal{A}}_g = \{ \tilde{A}_g \in \mathcal{A}_g : |\tilde{A}_g| = \tilde{\dot{r}}_g(n)  \}, \text{with}\,
    \tilde{\dot{r}}_g(n) \eqdef \max_{A_g \in \mathcal{A}_g} \{ |A_g| \}.
\end{equation}
Similarly $\mathcal{B}_g$ denotes the collections of sets satisfying \eqref{eq:14}, while $\tilde{\mathcal{B}}_g \subseteq \mathcal{B}_g$ comprises the maximal-cardinality subsets of size $\tilde{\ddot{r}}_g(n)$, i.e.:
\begin{equation}
    \label{eq:017}
    \tilde{\mathcal{B}}_g= \{ \tilde{B}_g \in \mathcal{B}_g : |\tilde{B}_g| = \tilde{\ddot{r}}_g(n) \}, \text{with}\,
    \tilde{\ddot{r}}_g(n) \eqdef \max_{B_g \in \mathcal{B}_g} \{ |B_g| \}.
\end{equation}
By accounting for \eqref{eq:016} and \eqref{eq:017}, it results that the volume of remote $n$-Gability and remote Pairability for a general two-colorable graph state are theoretically lower- and upper-bounded as follows, respectively:
\begin{align}
    \label{eq:018}
    r^{\ell_T}_g(n) & \eqdef \max \{\tilde{\dot{r}}_g(n) , \tilde{\ddot{r}}_g(n) \}  \leq r_g(n) \leq \lfloor \frac{N}{n} \rfloor, \\
    \label{eq:019}
    r^{\ell_T}_e & \eqdef \max \{\tilde{\dot{r}}_g(2) , \tilde{\ddot{r}}_g(2) \}  \leq r_g(2) \leq \lfloor \frac{N}{2} \rfloor .
\end{align}

\begin{lem}[\textbf{Mass $n_{\text{max}}$}]
    \label{lem:03}
    Given a $N$-qubit two-colorable graph state $\ket{G}$, with corresponding graph $G=(P_1, P_2, E)$, the highest mass $n_{\text{max}}$ of an extractable GHZ state among remote nodes satisfies the following inequality:
    \begin{align}
         \label{eq:16}
         n_{max}^{\ell} \eqdef \Delta (G) \leq n_{\text{max}} \leq \alpha(G)<N.
    \end{align}
    with $\Delta (G)$, $\alpha (G)$ given in Def.~\ref{def:05} and Def.~\ref{def:06}, respectively.
    \begin{IEEEproof}
        Let us denote with  $v_i \in G$, a  vertex characterized by $\deg(v_i)=\Delta(G)$. Performing Pauli-$y$ measurement on $v_i$ yields to the extraction of a GHZ state among all the neighbors of $v_i$. Hence $n_{\text{max}}$ is at least $\Delta(G)$. We define the constructive lower bound of $n_{\text{max}}$ to be $\Delta(G)$. The theoretical upper bound, $\alpha(G)$ directly follows from the remark after Def.~\ref{def:06}.
    \end{IEEEproof}
\end{lem}

\section{Algorithm}
\label{sec:4}

\begin{algorithm}[!t]
\caption{\textbf{\texttt{Remote Extraction}}{$(G, n)$}}
\label{alg:01} 
\begin{algorithmic}[1]
    \Require two-colorable graph  $G=(P_1, P_2, E)$
    \Ensure $n^{\ell}_{max},r^{\ell}_g(n),\mathcal{L}$
    
    \vskip 0.3em
    \Statex \hspace*{-\algorithmicindent} \rotatebox[origin=c]{-90}{\scalebox{0.8}{$\triangle$}}
    \textbf{\textit{Step 1: Approximating the maximum mass $n_{max}$ with the lower bound in \eqref{eq:16}}}
    \vskip 0.3em
    

    \For {$P$ in $(P_1, P_2)$}:
        \If{$\nexists v_i \in P$ with $N(v_i)=\bar{P}$}
        \Comment{\textit{Partition $P$ lacks star vertex}}
            \State $v_i \gets \arg \max_{v \in P} \deg(v)$ 
            \State $S, G \gets S\cup \{v_i\},  G \setminus \overline{N}(v_i)$ 
        \EndIf  \Comment{\textit{Updated $v_i$ into star set $S$ in partition $P$}}
    \EndFor \Comment{\textit{$S_{1(2)}$, $P_{1(2)}$, $V_{1(2)}$ given in \eqref{eq:06}-\eqref{eq:09}}}
    \State $n^{\ell}_{max} \gets \max\{\deg(v) \;\big|\; v \in (S_1 \cup S_2) \}$ 

    \vskip 0.3em
    \Statex \hspace*{-\algorithmicindent} \rotatebox[origin=c]{-90}{\scalebox{0.8}{$\triangle$}} 
    \textbf{\textit{Step 2: Computing the volume $r^{\ell}_g(n)$}}

    \vskip 0.3em
    \Statex \hspace*{-\algorithmicindent} \,\, \rotatebox[origin=c]{-90}{\scalebox{0.8}{$\triangle$}} \textit{Step 2.1: Drive initial $\tilde{r}_g(n) = |\tilde{A}_g|=\max\{|A_g|,|B_g|\}$ by random choosing $A_g, B_g$ in one partition}
    \vskip 0.3em
    
    \State $A_g \gets$ random.choice ($A_g \subseteq V_1:$ $A_g$ \text{satisfy} \eqref{eq:13}  \text{in} $G$)
    \State $B_g \gets$ random.choice ($B_g \subseteq V_1:$ $B_g$ \text{satisfy} \eqref{eq:14} \text{in} $G$)
    \If{$|B_g| > |A_g|$}
        \State $G, \tilde{A}_g \gets G \setminus \overline{N}_{\cap }(B_g), B_g$
    \Else
        \State $G, \tilde{A}_g \gets G, A_g$
    \EndIf

    \vskip 0.3em
    \Statex \hspace*{-\algorithmicindent} \,\,
    \rotatebox[origin=c]{-90}{\scalebox{0.8}{$\triangle$}}
    \textit{Step 2.2: Obtain $r^{\ell}_g(n)= |\hat{A}_g|$ after updating $\tilde{A}_g$ to $\hat{A}_g$}
    \vskip 0.3em
    \State $\texttt{A}, \texttt{\={A}2A} \gets$ \Call{ExpandA}{$G, n, \tilde{A}_g$}
    \While{$\texttt{A} \neq \emptyset$}
        \If{$\exists v_i \in \texttt{A} \text{ with } |\texttt{A}(v_i)| = 0$}  
            \State $\tilde{A}_g \gets \tilde{A}_g \cup \{ v_i \}$ \Comment{\textit{Add $v_i$ to $\tilde{A}_g$}}
        \Else
            \State Select any $v_i \in \texttt{A}$
            \If{$|\texttt{\={A}2A}(v_i)| = 0$}
                \State $G \gets G \setminus \big( \overline{N}({v_i}) \cap \overline{N}_{\cup}( \texttt{A}(v_i) ) \big)$
                \State $\tilde{A}_g \gets \tilde{A}_g \cup \{ v_i \}$
                \Comment{\textit{Add $v_i$ to $\tilde{A}_g$}}
            \ElsIf{$|\texttt{\={A}2A}(v_i)| = 1$}
                \State $v_j \gets \arg\limits_{v_i} |\texttt{\={A}2A}(v_i)| = 1$
                \State $G \gets G \setminus \big( \overline{N}({v_i}) \cap \overline{N}_{\cup}( \texttt{A}(v_i) ) \big)\setminus \{v_j\}$ 
                \State $\tilde{A}_g \gets (\tilde{A}_g \cup \{ v_i \}) \setminus \{ v_j \}$
                \Comment{\textit{Replace $v_j$ with $v_i$ in $\tilde{A}_g$}}
            \EndIf
        \EndIf
        \State $\texttt{A}, \texttt{\={A}2A} \gets$ \Call{ExpandA}{$G, n, \tilde{A}_g$}    
    \EndWhile
    \State $\hat{A}_g, r^{\ell}_g(n) \gets \tilde{A}_g, |\tilde{A}_g|$
    
    \vskip 0.3em
    \Statex \hspace*{-\algorithmicindent} \rotatebox[origin=c]{-90}{\scalebox{0.8}{$\triangle$}} 
    \textbf{\textit{Step 3: Identifying the location $\mathcal{L}$}}
    \vskip 0.3em
    \State $\mathcal{L} \gets \left\{ v_i \mapsto \overline{N}(v_i) \;\big|\; v_i \in \hat{A}_g \right\}$
    \State \Return $n^{\ell}_{max},r^{\ell}_g(n),\mathcal{L}$
\end{algorithmic}
\end{algorithm}

\floatname{algorithm}{Function}
\renewcommand{\thealgorithm}{1}
\begin{algorithm}[t]
\setstretch{1}    
\caption{\textbf{\texttt{ExpandA}}$( G, n, \tilde{A}_g )$}
\label{fun:x02}
    \begin{algorithmic}[1]
    \State $\bar{A} \gets \{ v_i \in V_i \setminus \tilde{A}_g : \left( |\overline{N}({v_i})| \geq (n-1) \right) \wedge \left( \overline{N}({v_i}) \not\subseteq \overline{N}_{\cup}({\tilde{A}_g}) \right)\}$ 
    \State $ \texttt{A}, \texttt{\texttt{B}2\texttt{A}}, \texttt{\={A}2A} \gets \emptyset, \emptyset, \emptyset$
    
    \For{$v_i \in \bar{A}$}
        \State $\texttt{\texttt{B}2\texttt{A}}(v_i) \gets \{ v_j \in \tilde{A}_g : \overline{N}(v_j) \cap \overline{N}(v_i) \neq \emptyset \}$
        \State $\texttt{\={A}2A}(v_i) \gets \{ v_j \in \tilde{A}_g : \overline{N}(v_j) \subseteq \overline{N}(v_i) \}$
    \EndFor
    
    \vskip 0.3em    
    \Statex \hspace*{-\algorithmicindent} \rotatebox[origin=c]{-90}{\scalebox{0.8}{$\triangle$}} \textit{Find $\texttt{A}(v_i)$ from $\texttt{\texttt{B}2\texttt{A}}(v_i)$, which can be combined with $v_i$ to form $B_g$, satisfying Lem.~\ref{lem:02}.}
    \vskip 0.3em

    \For{$v_i \in \texttt{\texttt{B}2\texttt{A}}$}
        \If{$|\texttt{\={A}2A}(v_i)| \leq 1$ \textbf{and} $\forall v_k \in \texttt{\texttt{B}2\texttt{A}}(v_i) \setminus \texttt{\={A}2A}(v_i), \{v_i, v_k\}$ satisfies \eqref{eq:14} in $G$}        
            \State $\texttt{A}(v_i) \gets \texttt{\texttt{B}2\texttt{A}}(v_i) \setminus \texttt{\={A}2A}(v_i)$ 
        \EndIf
    \EndFor

    \State \Return $\texttt{A}, \texttt{\={A}2A}$    
    \end{algorithmic}
\end{algorithm}

In the previous section, we  established theoretical bounds for the volume of remote $n$-Gability and remote Pairability, arising directly from the sufficient conditions established in Lemmas~\ref{lem:01} and~\ref{lem:02}. However, given the computational complexity of determining $\tilde{\dot{r}}_g(n),\tilde{\ddot{r}}_g(n)$, here we provide  constructive lower bounds for both remote $n$-Gability and remote Pairability, by designing an efficient algorithm for the extraction of remote entanglement resources. Then, in Sec.~\ref{sec:4.2}, we prove that such an algorithm exhibits a polynomial-time complexity.

\subsection{Algorithm Design}
\label{sec:4.1}
\begin{figure}[t]
    \centering
    \begin{subfigure}[b]{0.5\textwidth}
        \centering
        \includegraphics[width=0.95\textwidth]{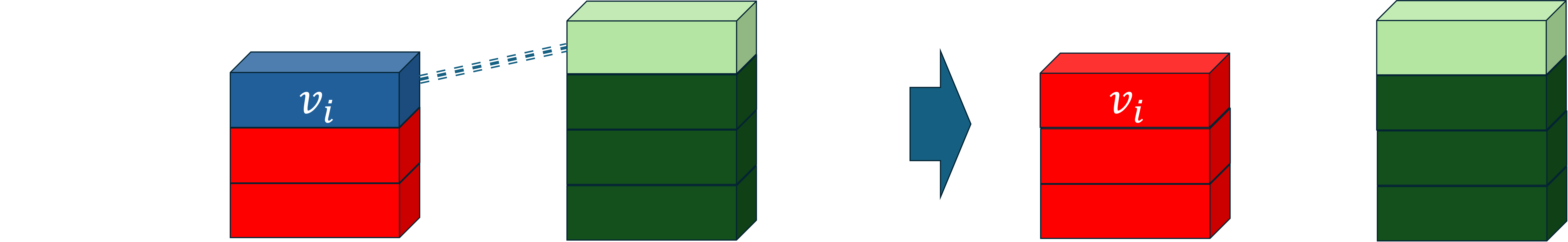}
        \subcaption{$|\texttt{A}(v_i)| = 0$}
        \label{fig:06.a}
    \end{subfigure}
    
    \vspace{6pt}

    \begin{subfigure}[b]{0.5\textwidth}
        \centering
        \includegraphics[width=0.95\textwidth]{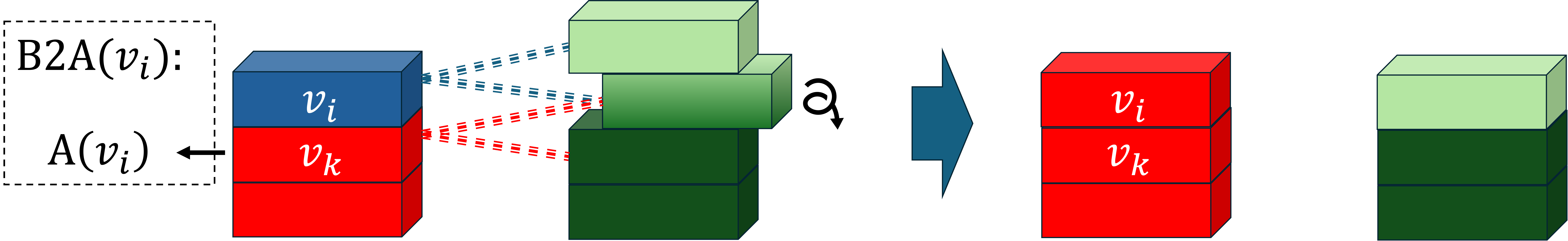}
        \subcaption{$|\texttt{A}(v_i)| \neq 0$ with $|\texttt{\={A}2A}(v_i)|= 0$}
        \label{fig:06.b}
    \end{subfigure}

    \vspace{6pt}
    
    \begin{subfigure}[b]{0.5\textwidth}
        \centering
        \includegraphics[width=0.95\textwidth]{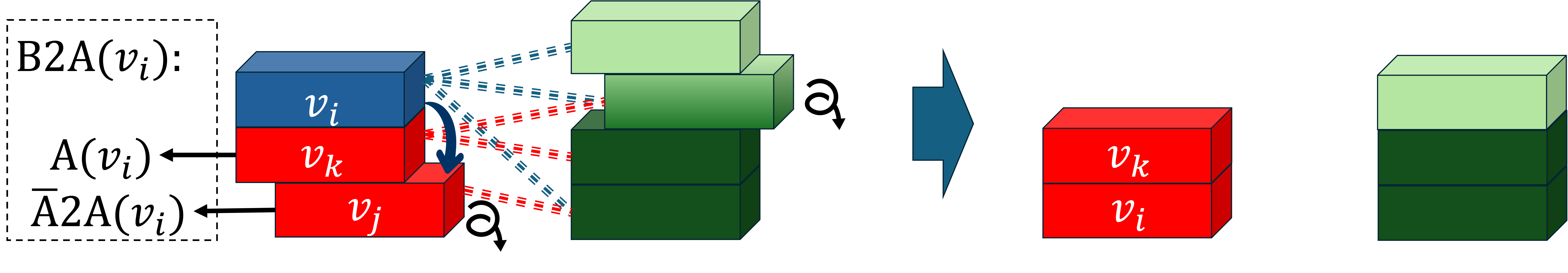}
        \subcaption{$|\texttt{A}(v_i)| \neq 0$ with $|\texttt{\={A}2A}(v_i)|= 1$}
        \label{fig:06.c}
    \end{subfigure}
    \caption{Pictorial representation of Step 2.2 in Algorithm~\ref{alg:01}. Dashed lines denote opposite remote sets (Def.~\ref{def:09}). \textbf{(a)} In the \texttt{\textsc{While}} loop, a vertex $v_i$ with $|\texttt{A}(v_i)| = 0$ (blue brick) is identified and added to $\tilde{A}_g$ (red bricks). \textbf{(b)} When no $v_i$ with $|\texttt{A}(v_i)| = 0$ exists, a vertex $v_i$ with $|\texttt{\={A}2A}(v_i)| = 0$ is found. After removing $\overline{N}({v_i}) \cap \overline{N}_{\cup}(A(v_i))$ (gradient green brick with "drop" icon), $v_i$ is added to $\tilde{A}_g$. \textbf{(c)} Otherwise, a vertex $v_i$ with $|\texttt{\={A}2A}(v_i)| = 1$ is found. After similar removal, $v_i$ is replaced by $\texttt{\={A}2A}(v_i)$, i.e., $v_j$ and added to $\tilde{A}_g$.}
    \label{fig:06}
    \hrulefill
\end{figure}

The proposed  algorithm for remote entanglement extraction is described in Algorithm \ref{alg:01}, organized into three steps:
\begin{itemize}
    \item[-] \textit{Step 1: Approximating the maximum mass $n_{max}$ of a remote GHZ with its lower bound in \eqref{eq:16}.} \\
    Let $G$ be the corresponding graph of $\ket{G}$. If $G$ does not have at least one star vertex in a certain partition $P$, then one vertex is updated as new star vertex in partition $P$ (Line 1-6). Then, by searching for the vertex with maximum degree in the star set $(S_1 \cup S_2)$, we approximate the maximum mass $n_{max}$ with its lower bound $n_{max}^\ell\eqdef \Delta(G)$, as proved in Lemma~\ref{lem:03}.
    
    \item[-] \textit{Step 2: Computing the volume $r^{\ell}_g(n)$ of extractable remote GHZ states, each of mass $n$.}\\
    This step is decomposed in two sequential sub-steps: \textit{Step 2.1}  deriving an initial estimation $\tilde{r}_g(n)$ of the volume; \textit{Step 2.2} iteratively refining $\tilde{r}_g(n)$ to obtain a more precise volume estimation $r^{\ell}_g(n)$, as detailed in the following.
    \begin{itemize}
    \item \textit{Step 2.1}: Two sets $A_g$ and $B_g$ are randomly chosen in $G$, by  satisfying equations \eqref{eq:13} and \eqref{eq:14}, respectively. By comparing the cardinality of these two set, an initial estimation $\tilde{r}_g(n)$ of the volume is provided: \begin{equation}
        \label{eq:17}
        \tilde{r}_g(n)=|\tilde{A}_g|\eqdef  \max\{|A_g|,|B_g|\}.
    \end{equation}
    
    \item \textit{Step 2.2}: It consists of an iterative refinement of $\tilde{A}_g$, through a stepwise expansion in lines 15 - 31, to obtain a more granular value $r_g^\ell(n)$ of the extractable volume, as illustrated pictorially in Fig.~\ref{fig:06}. This is achieved by calling the function ``ExpandA'', which iteratively (within the \texttt{\textsc{while}} loop, lines 16 - 31) scans all the vertices not in $\tilde{A}_g$ (constituting the $\overline{A}$ set in the function ``ExpandA'') to build an expanded set $\tilde{A}_g$ (lines 18 and 23) of higher cardinality. In particular, $\tilde{A}_g$ is updated either by directly adding $v_i \in \overline{A}$ to $\tilde{A}_g$ (as depicted in Fig.~\ref{fig:06.a} and~\ref{fig:06.b}), or by
    replacing  ``weaker'' (in terms of opposite remote-set)  vertices in the original  $\tilde{A}_g$ with vertices in $\overline{A}$ (as depicted in Fig.~\ref{fig:06.c}).
    At line 32,  the expanded $\tilde{A}_g$ is stored in  $\hat{A}_g$. The cardinality, $r^{\ell}_g(n)$, of the output $\hat{A}_g$  is the final volume for the remote $n$-Gability, i.e., by accounting for \eqref{eq:17}, it results:
    \end{itemize}
    \begin{equation}      
    \label{eq:18}
         r^{\ell}_g(n) =|\hat{A}_g| \geq \, |\tilde{A}_g| =\tilde{r}_g(n) \, \text{in} \, \eqref{eq:17}.
    \end{equation}
    Notably, $r^{\ell}_g(n)$ in \eqref{eq:18} serves as constructive lower bound for the volume of remote $n$-Gability.
    \item[-] \textit{Step 3: Identifying the identities of the nodes involved in the entangled resource.}\\
    The location $\mathcal{L}$ of the extracted entangled resources among remote nodes is determined by mapping each node $v_i$ in the set $\hat{A}_g$ to its opposite remote-set $\overline{N}(v_i)$. In other words, each extracted GHZ state is identified by $\{v_i, \overline{N}(v_i)\}$, with $v_i \in \hat{A}_g$.
\end{itemize}
 Clearly, Alg.~\ref{alg:01} also provides a constructive strategy for the remote Pairability, when $n=2$.

\begin{figure*}[t]
    \centering
    \begin{subfigure}[b]{\textwidth}
        \begin{subfigure}[t]{0.325\textwidth}
            \centering
            \includegraphics[width=\textwidth]{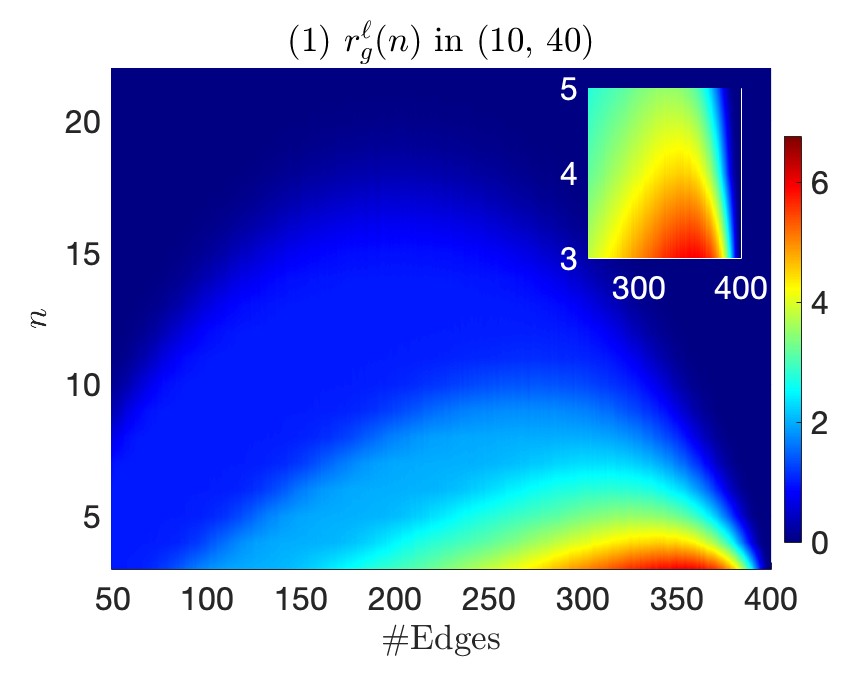}
            \label{fig:07.a}
        \end{subfigure}
        \begin{subfigure}[t]{0.325\textwidth}
            \centering
            \includegraphics[width=\textwidth]{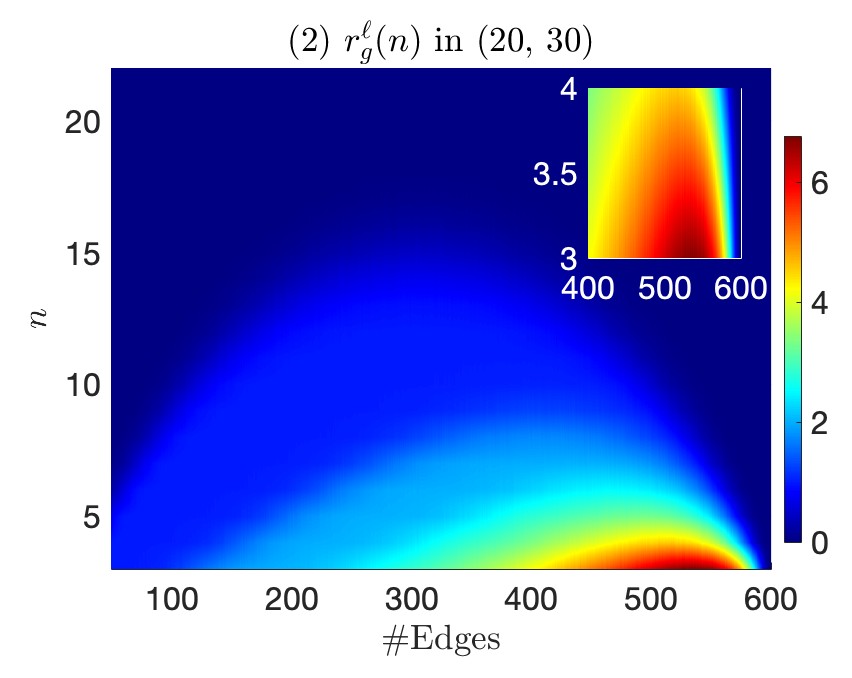}
            \label{fig:07.b}
        \end{subfigure}
        \begin{subfigure}[t]{0.325\textwidth}
            \centering
            \includegraphics[width=\textwidth]{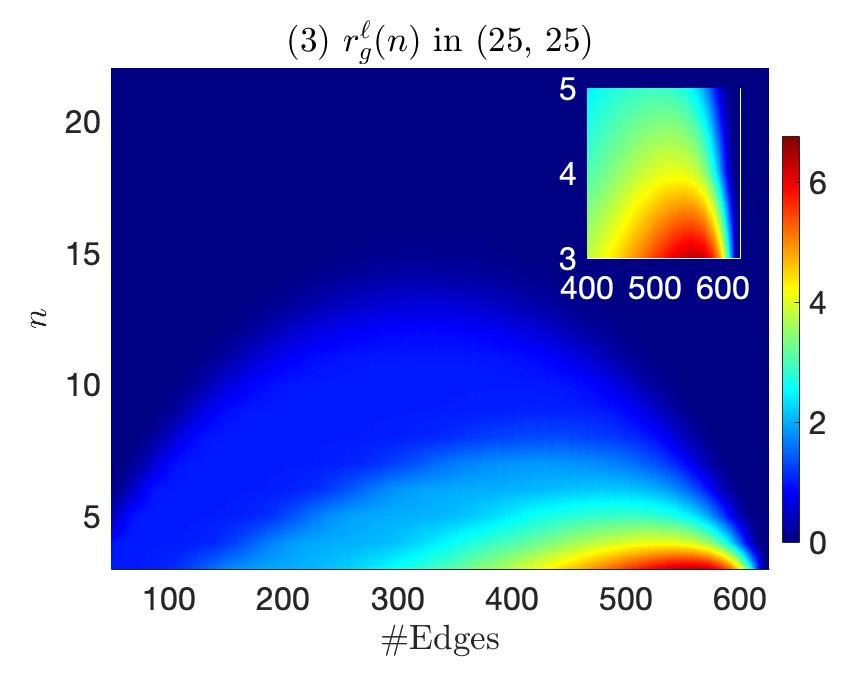}
            \label{fig:07.c}
        \end{subfigure}
    \end{subfigure}
    \setlength{\abovecaptionskip}{-0.35cm}  
    \caption{Remote $n$-Gability Performance Analysis: Average volume $r^{\ell}_g(n)$ for different  graph state partitions.}
    \label{fig:07}
    \hrulefill
\end{figure*}

\begin{figure}[t]
    \centering    
    \includegraphics[width=0.5\textwidth]{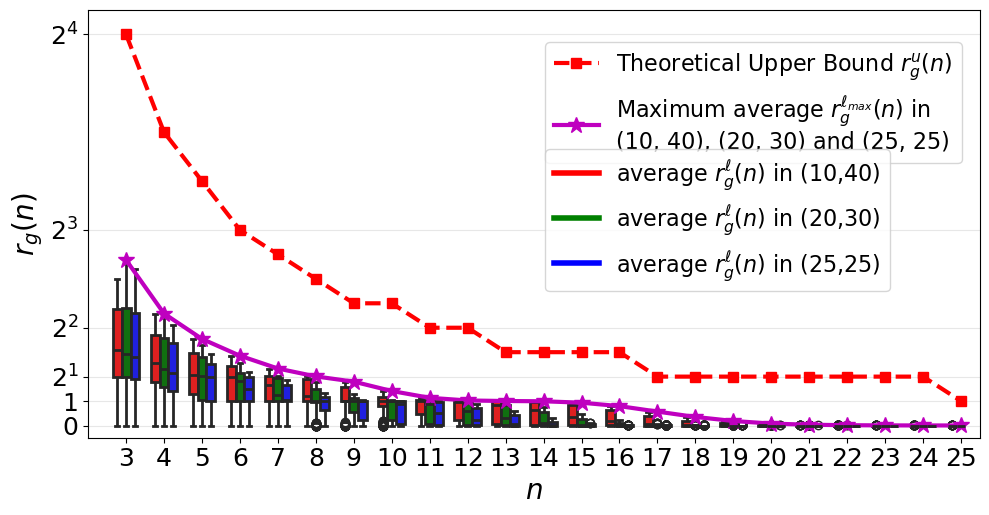}
    \caption{Remote $n$-Gability Performance Analysis: The average constructive lower bound $r^{\ell}_g(n)$ of the volume against the theoretical upper bound $r^{u}_g(n)$.}
    \label{fig:08}
    \hrulefill
\end{figure}

\subsection{Algorithm Complexity Analysis}
\label{sec:4.2}

Here, we analyze the complexity of Algorithm \ref{alg:01}. Let us assume, without any loss in generality, that the two-colorable graph $G=(P_1, P_2, E)$ is characterized by having $|P_1| \leq |P_2|$, with $P_1$ and $P_2$ defined in~\eqref{eq:08} and \eqref{eq:09}. 

\begin{theo}
\label{theo:01}
For any two-colorable graph state $\ket{G}$, with corresponding graph $G=(P_1, P_2,E)$, Algorithm~\ref{alg:01} determines:
\begin{itemize}
    \item a lower bound $r^{\ell}_g(n)$ of the remote $n$-Gability volume and the location of the extracted $n$-qubit GHZ states,
    \item a lower bound $r^{\ell}_e$ of the remote Pairability volume and the location of the extracted EPR pairs,
\end{itemize}
within a time complexity of $O(|P_1|^3*|P_2|)$.
    \begin{IEEEproof}
        Please refer to App.~\ref{app:theo:01}.   
    \end{IEEEproof}
\end{theo}

\begin{cor}   
\label{cor:04}
For any two-colorable graph state $\ket{G}$, with corresponding graph $G=(P_1, P_2,E)$, Algorithm~\ref{alg:01} determines the maximum mass $n_{max}$ of an extractable GHZ state among remote nodes with time complexity $O(|P_1|*|P_2|)$.
    \begin{IEEEproof}
        Please refer to App.~\ref{app:theo:01}.   
    \end{IEEEproof}
\end{cor}

\begin{remark}
    Our algorithm leverages graph theory tools and uses only single-qubit Clifford operations, single-qubit Pauli measurement and classical communications. Theo.~\ref{theo:01} shows that Alg.~\ref{alg:01} can  determine the volume and location of the vertices involved in the extracted entangled resources in polynomial-time. Furthermore, the maximum mass can also be obtained in polynomial time.\\
    It is worthwhile to emphasize that Alg.~\ref{alg:01} computes a lower bound on the volume of the extractable entangled resources.  
 By summarizing, the volume of the remote $n$-Gability and remote Pairability of a general two-colorable graph state are constructively bounded as follows:
\begin{align}
    \label{eq:30}
    r^{\ell}_g(n) \; \text{given in} \; \eqref{eq:18} & \leq r_g(n) \leq \lfloor \frac{N}{n}  \rfloor, \\
    \label{eq:31}
    r^\ell_e =r^{\ell}_g(2) \; \text{given in} \; \eqref{eq:18} &\leq r_e \leq \lfloor \frac{N}{2} \rfloor.
\end{align}\color{black}
This result is remarkable, since, as stated in Sec.~\ref{sec:1}, no known algorithm -- even with exponential-time complexity -- guarantees an exact solution for all graph state structures. The underlying theoretical problem is indeed NP-complete. 
\end{remark}

\section{Performance Evaluation}
\label{sec:5}

Here, we conduct a performance analysis, by considering  both general two-colorable graph states and representative Internet-inspired artificial topologies. 

As discussed in Sec.~\ref{sec:4.2}, no exact procedure currently exists to determine the optimal solution of the Remote-VM problem. Therefore, our analysis is conducted under a worst-case scenario, in which the obtained results are compared against  theoretical upper bounds. 
These upper bounds are inherently very conservative, being static and independent of the specific graph state instance. In contrast, the lower bounds computed by Algorithm~\ref{alg:01} are not only constructive but are able to dynamically adapt to different graph structures. To rigorously account for structural heterogeneity across graph instances, the results are averages over 1,000 independently generated graph instances.

\subsection{General Two-colorable Graph State Performance}
\label{sec:5.1}

We first conducted a comprehensive evaluation of general two-colorable graph states under various graph structures. An explicit setup process is provided, along with a detailed analysis of remote pairability and remote $n$-Gability.

\subsubsection{\textbf{Setup}}
\label{sec:5.1.1}

We evaluate the extractable values against different bipartite graph structures by randomly varying the number of edges $m$, while keeping the total number of nodes constant and equal to $50$. This allows for a fair comparison across various graph instances.
Furthermore, for the sake of generality, we distribute the nodes in different ways: one approach allocates nodes unequally across partitions, while the other ensures an equal number of nodes in each partition. 
Specifically, we consider graphs with partitions ($P_1, P_2$) of sizes (10, 40), (20, 30), (25, 25), respectively.
We then randomly distribute the $m$ edges between the two partitions, thereby varying the graph structure. For being adherent to the definition of bipartite graph state, the number of edges in the corresponding graph has to satisfy the following inequality:
\begin{equation}
    \label{eq:19}
    (|P_1|+|P_2|-1) \leq m \leq |P_1|*|P_2|.
\end{equation}

\begin{figure}[t]
    \centering
    \includegraphics[width=0.5\textwidth]{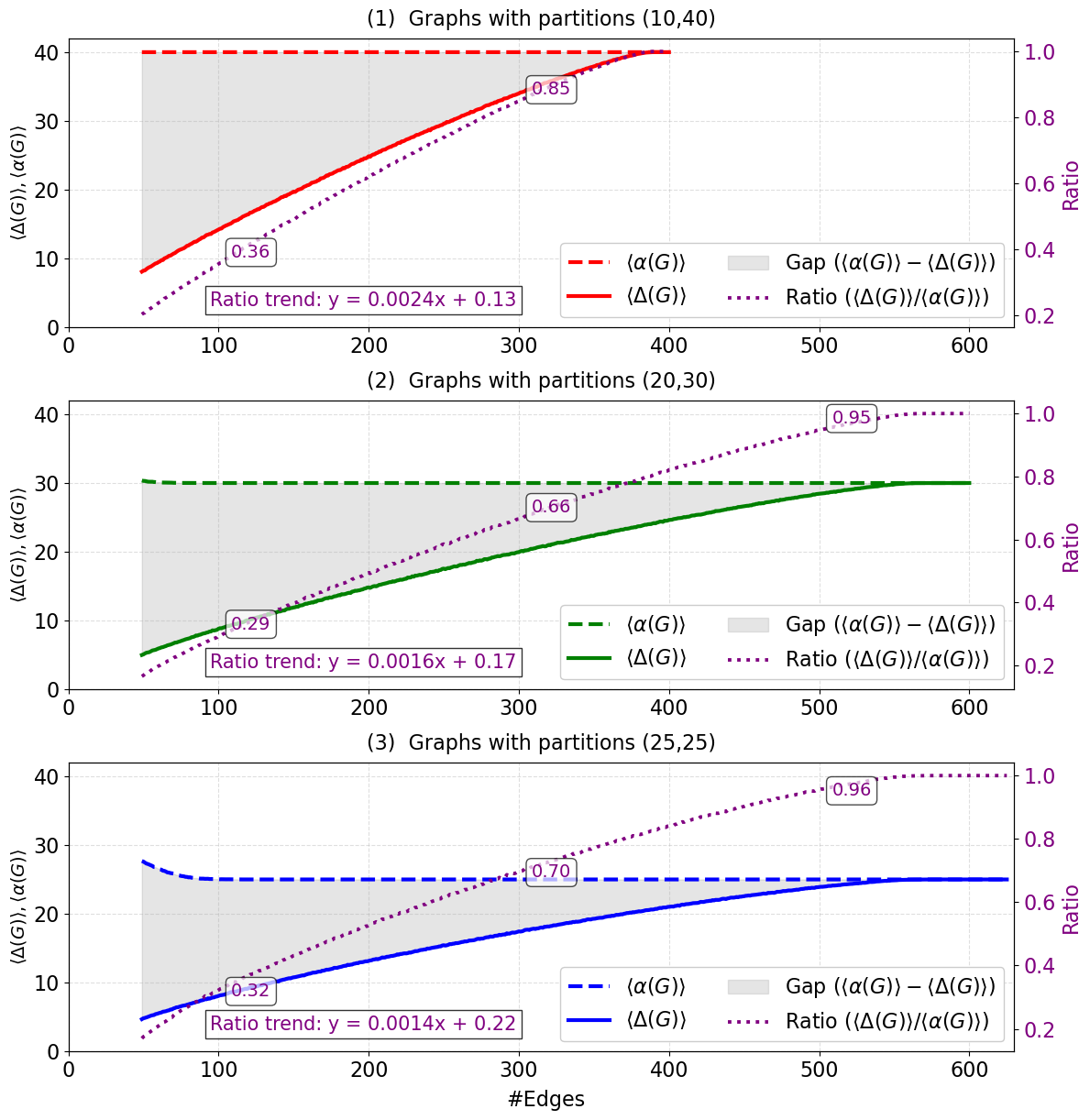}
    \caption{Maximum Mass $n_{max}$ of Remote $n$-Gability: Average lower- $\Delta(G)$ and upper $\alpha(G)$ bounds, for different graph partitions (10, 40), (20, 30), (25, 25), respectively.}
    \label{fig:09}
    \hrulefill
\end{figure}

\subsubsection{\textbf{Remote $n$-Gability Performance Analysis}}
\label{sec:5.1.2}

To evaluate the volume for remote Gability, we compute via Algorithm~\ref{alg:01} the average $r^{\ell}_g(n)$ in \eqref{eq:18}, which serves as constructive lower bound.
Fig.~\ref{fig:07} validates the $n$-Gability volume, for each configuration of bipartite graph state and against not only the number $m$ of edges but also against the mass $n$ of the extracted GHZ states. 
As shown in Fig.~\ref{fig:07}, our approach  generally enables the extraction of at least one GHZ state with a mass ranging from 3 to 17 among remote nodes. This implies that for a given graph state, one can typically extract a GHZ state of significant size among distant parties. 
Notably, for $\ket{GHZ}_3$ states, we observe that the volume $r^{\ell}_g(3)$ surpasses 6 across all considered partition configurations. 
This indicates that our method efficiently facilitates the formation of small-scale GHZ states (i.e., small subnets) that can be exploited by entanglement-based protocols.
Furthermore, Fig.~\ref{fig:08} compares the theoretical upper bound with the maximum average extractable volume $r^{\ell}_g(n)$ for $n \in [3,25]$ in graph states with partitions (10, 40), (20, 30), and (25, 25). 
As $n$ increases, the gap between the bounds narrows. Given that the theoretical upper bound is inherently conservative -- being static and instance-agnostic -- this trend indirectly confirms the tightness and efficacy of our constructive lower bounds.

Furthermore, we stress that existing studies focus on maximizing the mass of a single GHZ state, by limiting the volume to be equal to one. For a graph state $\ket{G}$ with bounded rank-width, in~\cite{Oum-08} a poly-time algorithm determines whether a GHZ state can be extracted using local Clifford operations and Z-measurements, providing the required operation sequence. Similarly,~\cite{JonHahTch-24} demonstrates the extraction of GHZ states with masses from 4 to 11 starting from linear cluster states of up to 19 qubits on the IBMQ Montreal quantum device.
By accounting for the above, our results not only demonstrate the extraction of GHZ states with significantly larger
masses ranging from 3 to 17, but also ensure the extraction of a considerable volume of 3-qubit $\ket{GHZ}$ states. This showcases the versatility of our method, enabling both large and small-scale GHZ states, and providing a scalable and efficient approach for quantum networks.

\textit{\textbf{Maximum Mass $n_{max}$ Performance Analysis}}:
\label{sec:5.1.2.1}
To evaluate the extractable maximum mass for remote Gability, we compute the average lower bound $\Delta(G)$ and corresponding theoretical upper bound $\alpha(G)$, given in~\eqref{eq:16}. Fig.~\ref{fig:09} validates the mass analysis, for each configuration of bipartite graph state and against the edge number $m$.
Specifically, both $\Delta(G)$ and $\alpha(G)$ are affected by the partition ratio ($|P_1|:|P_2|$). And as the partition ratio approaches 1, $\alpha(G)$ decreases. Notably, $\alpha(G)$ remains largely unaffected by the number of edges $m$, whereas for each bipartite graph configuration, $\Delta(G)$ increases with $m$ and eventually converges to $\alpha(G)$. Intuitively, a more balanced partition ($|P_1|:|P_2|$ closer to 1) results in a smaller gap between $\Delta(G)$ and $\alpha(G)$. To illustrate this, we also plot the trend of the ratio in Fig.~\ref{fig:09}.

\begin{figure}
    \centering
    \includegraphics[width=0.95\linewidth]{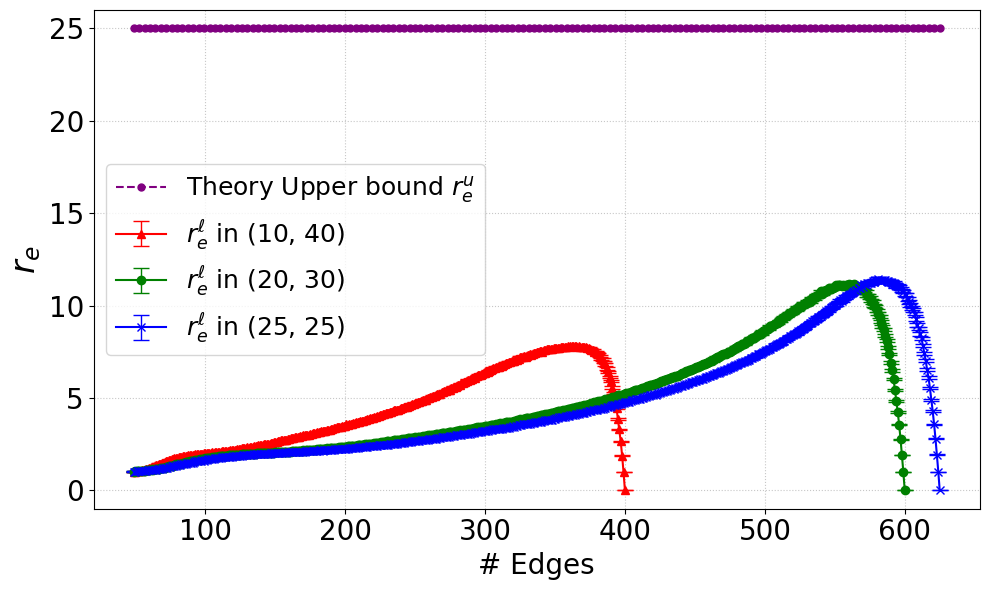}
    \caption{Remote Pairability Performance Analysis I: $95\%$ confidence interval of the volume lower-bound $r^{\ell}_e$, with partitions (10, 40), (20, 30), (25, 25), respectively. The figure also shows the theoretical upper bound  $r^{u}_e$, for the remote Pairability volume.}
    \label{fig:10}
\end{figure}

\begin{figure*}[t]
    \centering
    \includegraphics[width=\textwidth]{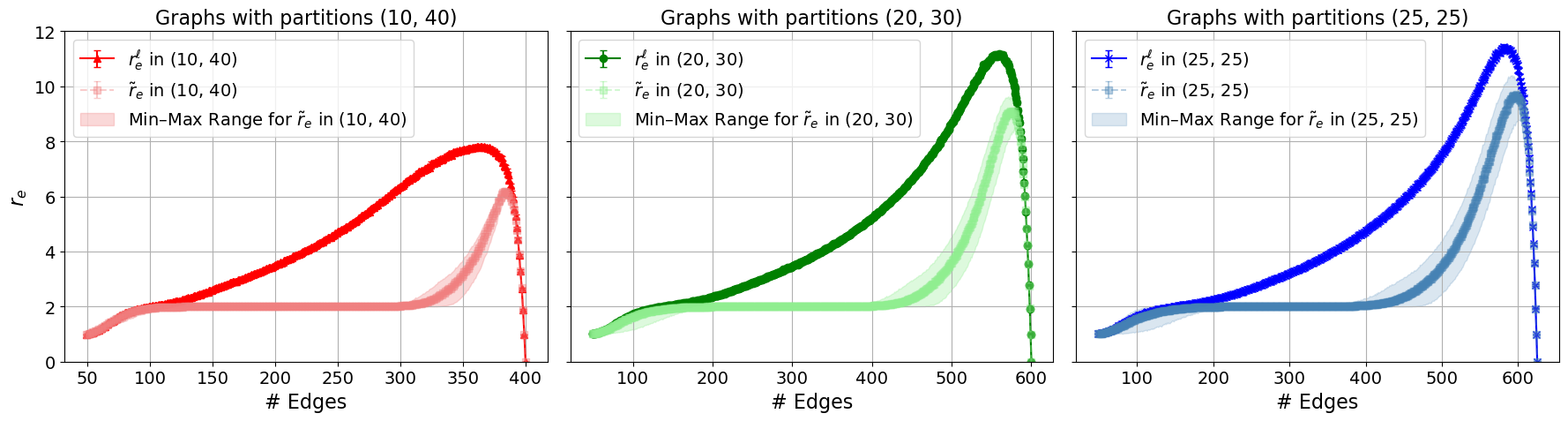}
    \caption{Remote Pairability Performance Analysis II:  $95\%$ confidence interval for $r^{\ell}_e$ against the preliminary estimation $\tilde{r}_e$ in \eqref{eq:17}, with volumes from Step 2.1 of Alg.\ref{alg:01}. The figure also shows min-max range of $\tilde{r}_e$ for each configuration of graph state. 
    It is noteworthy that across all partition configurations, the value of $\tilde{r}_e$ consistently lies below that of $r^{\ell}_e$, which validates the effectiveness of Alg. \ref{alg:01}. A comparison of the three subfigures reveals that the behavioral pattern of the metrics varies significantly with the partition size. This is observable in the differing value ranges on the y-axes and the distinct trends as the number of edges (x-axis) increases.
    }
    \label{fig:x10}
    \hrulefill
\end{figure*}

\begin{figure*}[t]
    \centering    \includegraphics[width=\textwidth]{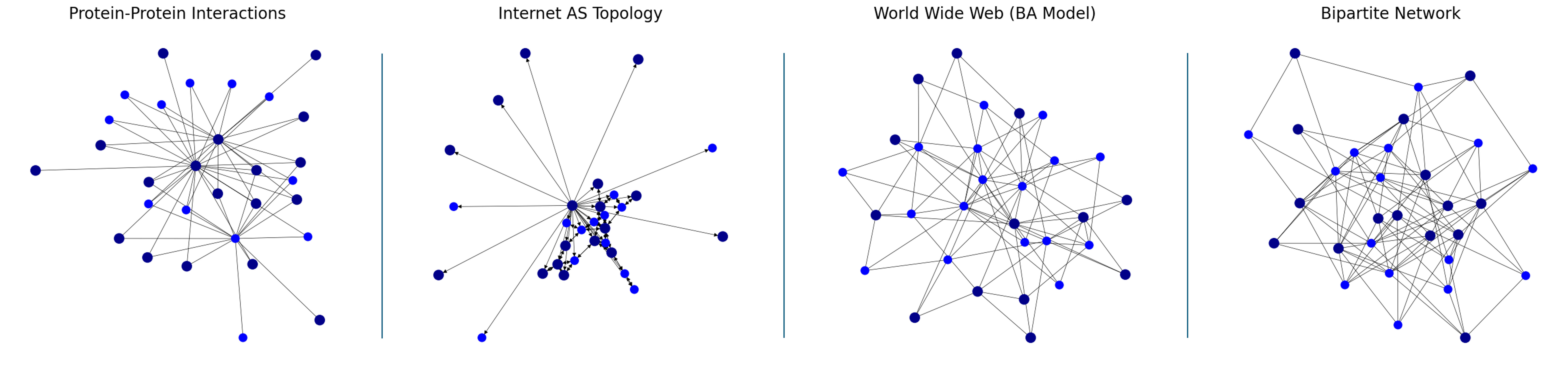}
    \caption{Complex artificial topologies used in evaluation. The graphs were generated using the NetworkX library.}
    \label{fig:x11}
    \hrulefill
\end{figure*}

\begin{figure*}[t]
    \centering    \includegraphics[width=0.8\textwidth]{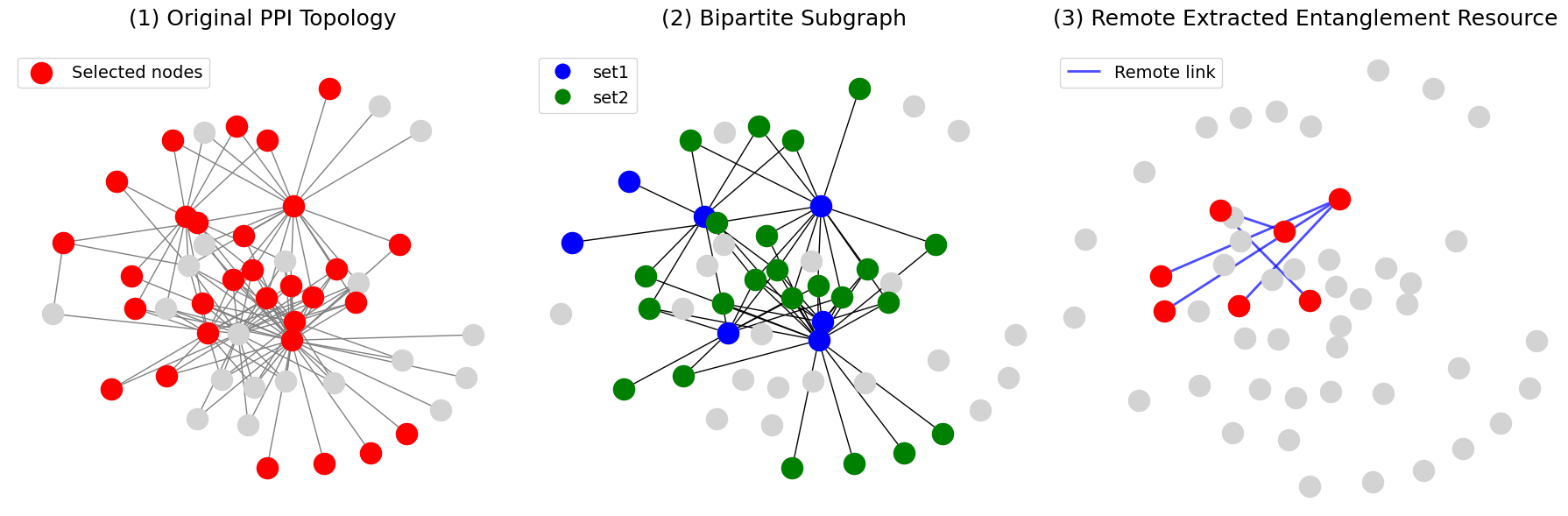}
    \caption{Pictorial illustration for Remote extraction from general graph state. 
    (1) The sample Protein-Protein Interactions (PPI) topology is generated by NetworkX library. (2) The extract bipartite subgraphs from PPI topology with fixed number of nodes. (3) The remote extracted entanglement resource, i.e., a $4$-qubit GHZ and a $3$-qubit GHZ, can be obtained by Alg.~\ref{alg:01}.}
    \label{fig:x12}
    \hrulefill
\end{figure*}

\begin{figure}[t]
    \centering    \includegraphics[width=0.5\textwidth]{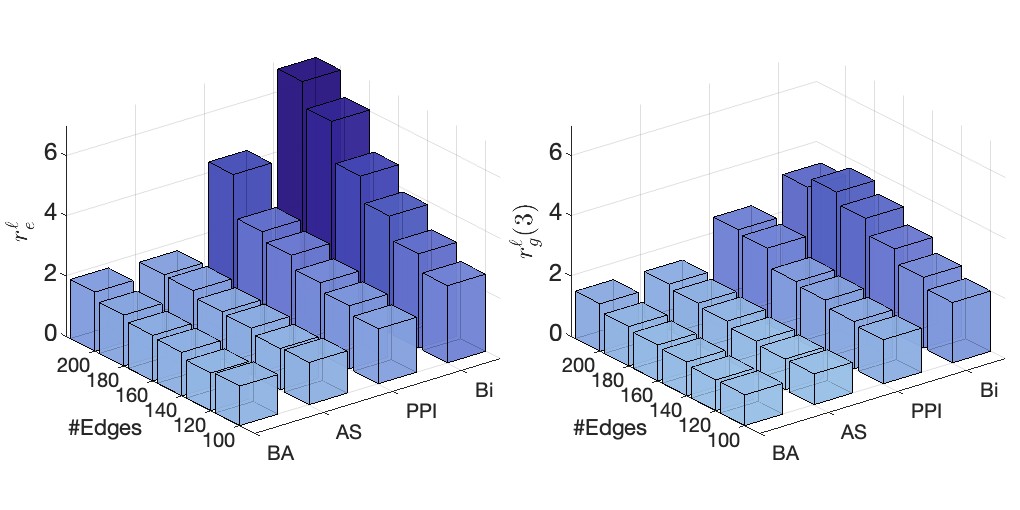}
    \caption{Remote Pairability and Remote Gability Performance Analysis: Average extractable volume $r^{\ell}_e$, $r^{\ell}_g(3)$ in general 50-qubit graph state with BA-model, AS Internet, Protein-Protein Interactions, and Bipartite network topology, respectively.}
    \label{fig:13}
    \hrulefill
\end{figure}

\subsubsection{\textbf{Remote Pairability Performance Analysis}}
\label{sec:5.1.3}
To evaluate the remote Pairability for general two-colorable graph states, we compute the $95\%$ confidence interval of the constructive lower-bound $r^{\ell}_g(2)$, given in \eqref{eq:18}. 
Specifically, in Fig.~\ref{fig:10}, we present both theoretical upper bound and our constructive lower bound. We observe an intriguing contrast in the performance of $r^{\ell}_e$ across the three different partitions size of graph states. Specifically, the (25, 25) configuration exhibits the highest $r^{\ell}_e$ values (11-12), indicating more pronounced extractable capabilities in balanced graph structures. 

To demonstrate the effectiveness of our algorithm, in Fig.~\ref{fig:x10} we also compare the preliminary estimation $\tilde{r}_e=\tilde{r}_g(2)$ in \eqref{eq:17} of the remote Pairability volume, provided in the Step 2.1 of Alg.~\ref{alg:01}, with the final constructive lower bound $r^{\ell}_e$. The figure shows that $r^{\ell}_e$ is significantly higher that the initial estimation $\tilde{r}_e$, for all the three partition configurations. 
The aformentioned trend is observed by excluding, as expected, the sparse graph regime and the high density regime.


In this context, we further note that a direct comparison between the bounds established in this work and those reported in the existing literature may not be entirely appropriate, as -- to the best of our knowledge -- this is the first study to explicitly address remote pairability, in contrast to conventional (or vanilla) pairability. Existing studies, such as \cite{Zha-25} and \cite{BraShaSze-22}, focus primarily on determining whether subsets of Bell pairs can be extracted from specific topologies -- such as rings, linear chains, and trees,  by allowing the inclusion of adjacent nodes in this extraction process. For instance, \cite{Zha-25} derives conditions for extracting two EPR pairs from such structures, without enforcing the remoteness constraint. Similarly, \cite{BraShaSze-22} identifies a 2-pairable 10-qubit wheel graph state via exhaustive numerical evaluation of all possible Pauli measurement sequences, yet again without ensuring remoteness. In contrast, for the same wheel graph structure, our approach guarantees the extraction of at least one remote EPR pair through a constructive procedure.

\subsection{General Graph State Performance}
\label{sec:5.2}
In the following, we evaluate the extractable volumes $r^{\ell}_e$ and $r^{\ell}_g(n)$ for general graph states across different graph structures. 
To enhance the generality of our analysis we consider four representative Internet-inspired topologies: World Wide Web, using the Barabási–Albert (BA) model, Autonomous System (AS) Internet, Protein-Protein Interaction networks, and Bipartite network topology, as illustrated in Fig.~\ref{fig:x11}. These topologies may capture structural features envisioned in future quantum networks. For each of these topologies, we conduct evaluations on remote $n$-Gability and remote Pairability.

For a fair and consistent evaluation across topologies, we fix the node count at 50 and use the NetworkX library to generate graph instances with varying numbers of edges $m$. Since the original Internet-inspired topologies are not necessarily bipartite, Algorithm~\ref{alg:01} cannot be directly applied to compute their extractable volumes. To address this, we extract bipartite subgraphs from each 50-node topology, by selecting 30 nodes that form a bipartite graph. An example for the Protein-Protein Interactions (PPI) topology is shown in Fig.~\ref{fig:x12}. To ensure statistical reliability, we performed 100 experiments to generate random graphs for each edge number scenario. This allows for a fair comparison across topologies under consistent structural conditions.

To evaluate the remote Pairability for general graph states, we compute the average lower-bound $r^{\ell}_e$, in Fig.~\ref{fig:13}. We observe that the volume can be successfully determined for each type of Internet topology. As the number of edges increases -- i.e., as the network topology becomes denser -- the extractable volume exhibits an approximately linear growth trend.

For remote $n$-Gability, we evaluate the extractable volume for 3-qubit GHZ states, i.e., $r^{\ell}_g(3)$, as a representative case. The results are shown in Fig.~\ref{fig:13}. Similar to the remote pairability case, we are able to determine the extractable volume for all tested Internet topologies. In general, the extractable volume increases roughly linearly with edge density. However, we also observe a performance drop in extremely dense bipartite network structures, where the extractable volume does not continue to increase and may slightly decline.

 
\section{Conclusion and Discussion}
\label{sec:6}
In this work, we investigated remote $n$-Gability and remote Pairability for general graph states, quantifying how many
GHZ states (of any size $n$) and EPR pairs can be concurrently extracted only between non-adjacent nodes in the artificial topology. To the best of our knowledge, this is the first study assessing the remote extraction abilities of multipartite entanglement states. Specifically, we formulated the Remote-VM problem and proposed a constructive polynomial-time algorithm to address it.
We derived both constructive lower bounds and theoretical upper bounds of the extractable volume, identified the location of each involved vertex, and characterized the range of the maximum mass, extractable among remote nodes.

In the following, we discuss key implications of the proposed framework and its underlying rationale, by outlining potential and promising directions for future research.

\textbf{Counting problem:} As discussed in the Introduction, although the Remote-VM problem shares conceptual foundations with the $\#IS$ counting problem, it introduces uniquely quantum challenges.
Unlike the static nature of $\#IS$, entanglement extraction is an inherently dynamic process, since each local measurement induces global changes in the graph state, by making both the choice and order of operations critical. This temporal dependence, which has no-analogue in classical counting, introduces a fundamental challenge, as a single misstep can highly influence the target resource. Furthermore, practical overheads connected to the extraction process, such as the consumption of measured qubits, further widen the gap between the theoretical counting maximum and what is operationally achievable. These fundamental differences imply that $\#IS$ serves as a theoretical upper bound for extractable entanglement, while our framework establishes a constructive lower bound. Bridging this gap, by developing operational strategies that more closely approach the theoretical limit, represents a compelling direction for future research at the intersection of graph theory and quantum information.

\textbf{Extraction Methods:}  
As aforementioned, the quantitative bounds presented in this work are established under the stringent constraints of the Remote-VM scenario, which fundamentally differs from the objectives pursued in vanilla extraction literature. Consequently, a direct numerical comparison would be neither expressive nor informative, since the state-of-the-art algorithms are not conceived and, in practice, cannot be optimized to satisfy the stringent remoteness constraint.\\ 
This difference in scope inherently leads to divergent outcomes, even when applied on the same graph. For instance, in the wheel graph state, our method yields one fewer EPR pair than that reported in~\cite{BraShaSze-22}, since their notion of ``pairability'' does not enforce the remoteness constraint. In other situations, however, our proposal exhibits superior performance, especially when existing methods primarily target simplified topologies such as linear chains or trees.\\ 
To better substantiate the above statement, consider a $6$-qubit butterfly topology, where our Algorithm~1 successfully extracts two remote EPR pairs. In contrast, adapting existing extraction methods to this butterfly topology would require a preliminary yet challenging graph simplification step.
Specifically, standard simplification techniques such as graph foliage~\cite{Zha-25} fail to reduce the graph, while deleting a bottleneck node to linearize the structure yields only one remote EPR pair in~\cite{Zha-25}.\\ This qualitative comparison demonstrates that the higher yields reported in prior works are achieved by relaxing the remoteness constraint, thereby highlighting the fundamental trade-off in our problem formulation.

\textbf{$N$-colorable Graph State:} The proposed framework established for two-colorable graph states provides a solid foundation for arbitrary graph states, with the associated conversion overhead representing an interesting direction for future research. Indeed, since a graph is two-colorable if and only if it contains no odd cycles~\cite{HeiDurEis-06}, the aforementioned conversion requires the elimination or neutralization of such cycles. Two main approaches can be employed for this purpose. The ancillary-based conversion neutralizes odd cycles by inserting auxiliary qubits along selected edges~\cite{ZahRez-11}, thereby ensuring bipartiteness at the cost of -- at most -- a linear blow-up in the total number of qubits. Conversely, measurement-based extraction removes odd cycles via vertex deletion through Pauli-$Z$ measurements~\cite{HeiDurEis-06,HeiEisBri-04},  
yielding a bipartite vertex-minor of the original graph but at the cost (in general) of reduced entanglement.
These considerations position two-colorable graph states as a fundamental and representative framework for arbitrary graph states. A promising direction for future work is the design of adaptive hybrid strategies that dynamically choose between ancillary introduction and vertex-minor extraction based on the specific topology and resource constraints.

\textbf{Noisy Scenarios:} The extension of our framework to noisy scenarios opens several promising research directions grounded in the present work. Notably, the stabilizer formalism underlying our proposal inherently suggests a degree of robustness against Pauli-noise. Indeed, as shown in~\cite{DurBri-04}, any Pauli-noise map acting on a graph state can be equivalently rewritten as a noise map containing only products of $Z$ and identity operators, owing to the stabilizer structure of graph states. Crucially, the Remote-VM approach's core operation -- Pauli-$X$ measurement on each star vertex -- is itself represented, within the stabilizer formalism, by the application of $Z$-type byproduct operators on its neighbours. Consequently, the same algebraic structure governs both the measurement-based transformations and the noise processes, ensuring that the core mechanism of the Remote-VM approach remains structurally robust under Pauli-noise. \\
A compelling future research direction is the investigation of noise-adaptive extensions of our framework. Specifically, we envision that the choice of star vertices could be guided by device-specific error profiles, for instance, by prioritizing nodes less affected by Pauli-$X$ errors, without altering the core extraction logic. Moreover, leveraging the stabilizer structure of graph states can guide the design of integrated noise-aware vertex selection and error-mitigation strategies, leading to a unified and resilient framework for remote entanglement extraction.

\section*{Acknowledgment}
\noindent We thank Dr. Jessica Illiano for her insightful suggestions on an earlier version of this manuscript.

\begin{appendices}

\section{PROOF OF LEMMA~\ref{lem:01}}
\label{app:lem:01}


We assume that equation~\eqref{eq:13} holds, and we must prove that $\dot{r}_g(n)$ GHZ states with each GHZ involving at least $n$ qubits can be extracted from the graph state $\ket{G}$. Let us assume, without loss of generality, $A_g \subseteq V_1 \subseteq P_1$ and let us follow the labeling given in \eqref{eq:08} and \eqref{eq:09}. Additionally, in the following, we denote with $N^i \eqdef N(v^i_1)$ and $\overline{N}^i \eqdef \overline{N}(v^i_1)$ the set of neighbors and the set of opposite remote nodes for node $v_1^i$ in the original graph $G$, respectively. Conversely, we use $N(v^i_1)$ and $\overline{N}(v^i_1)$ for denoting the ``current'' identities of the nodes belonging to the respective sets during the manipulation of the graph. The proof constructively follows by performing the following four tasks. In a nutshell, 
the first two tasks remove irrelevant vertices which will not be linked by a GHZ state. The third task interconnects each vertex in $A_g$ with its opposite remote set, with the exception of an arbitrary vertex. Finally, the last task interconnects also such a vertex with its opposite remote set and removes extra links among the nodes in $A_g$, as detailed in the following.
\begin{itemize}
    \item[i)] Pauli-$Z$ measurements on the qubits corresponding to the vertices in $V_1 \setminus A_g$ plus all the start vertices in $S_1$ except one vertex, say $s_1^1$.
    \item[ii)] Pauli-$Z$ measurements on the qubits corresponding to the vertices in $V_2 \setminus \overline{N}_{\cup}(A_g)$
    plus all the start vertices in $S_2$ except one vertex, say $s_2^1$.\\
    These two tasks are equivalent to remove irrelevant vertices, which will not be linked by a GHZ state, with the exception of the two additional vertices, namely, $s^1_1$ and $s^1_2$. Thus, the former two tasks yield to the graph:
    \begin{small}
    \begin{equation}
        \label{eq:20}
        G' = G - \left( P_1 \setminus \left( A_g \cup \{s^1_1\} \right) \right) - \left( P_2 \setminus \left( \overline{N}_{\cup}(A_g)   \right) \setminus \{ s^1_2 \} \right).
    \end{equation}
    \end{small}

    \item[iii)] Pauli-$X$ measurement on the selected star vertex $s_2^1$ with the arbitrary neighbor $k_0 \in A_g$, denoted as $v_1^1$ for the sake of simplicity. 
    Thus, the third task yields the graph: 
    \begin{equation}
        \label{eq:21}
        G'' = \tau_{v^1_1} \left( \tau_{s^1_2}\big(\tau_{v^1_1}(G')\big)-s^1_2 \right).
    \end{equation}
    
    \item[iv)] Pauli-$X$ measurement on the star vertex $s_1^1$ by choosing again $v_1^1$ as the arbitrary neighbor $k_0$ (which belongs now to $N(s_1^1)$ as a consequence of the first Pauli-$X$ measurement).
    Thus, the forth task yields the graph:
    \begin{equation}
        \label{eq:22}
        G''' = \tau_{v^1_1} \left( \tau_{s^1_1}\big(\tau_{v^1_1}(G'')\big)-s^1_1 \right).
    \end{equation}
\end{itemize}   

From~\eqref{eq:22}, we have that, in the final graph, each node $v^i_1 \in A_g$ is connected with and only with all the nodes in the original opposite remote set $\overline{N}^i$. Hence, by considering the subgraph induced by the vertices $\{ v^i_1 \} \cup \overline{N}^i$, such a subgraph is a star subgraph with $v^i_1$ acting as star vertex, and each of these $\dot{r}_g(n) = |A_g|$ subgraphs is disconnected -- i.e., disjoint -- from the others subgraphs. Thus, the thesis follows.


\section{PROOF OF THEOREM~\ref{theo:01}}
\label{app:theo:01}

\textit{(1) Complexity for determining $n^{\ell}_{max}$}

The procedure (Lines 1-6) requires $O(|P_1|*|P_2|)$ time complexity to ensure each partition contains at least one star vertex. Then, the algorithm computes the lower bound of maximum mass $n_{mass}$, by evaluating the vertex degrees within the star sets $S_1$ and $S_2$, i.e., $n_{mass}^\ell=\max\{\deg(v) \;\big|\; v \in (S_1 \cup S_2)\}$. In fact, by definition, this is equal to $\Delta(G)$, namely the maximum degree of the graph. Accordingly, the procedure maintains an overall time complexity of $O(|P_1|*|P_2|)$.

\textit{(2) Complexity for determining $r^{\ell}_g(n)$ and the location of the involved vertices.}

\textit{Lines 8-14:} we firstly generate a random  permutation of $V_1$, which exhibits a time-complexity of  $O(|P_1|)$. Then the procedure constructs $A_g, B_g$, via sequential checks, which require a time complexity of $O(|P_1|^2*|P_2|)$ in the worst case. Then, if the cardinality of $B_g$ is larger than $A_g$, we remove the intersection of the opposite remote sets of $B_g$, namely $\overline{N}_{\cap}(B_g)$, from the vertex set. This exhibits a time complexity of  $O(|P_1|*|P_2|)$, again in the worst case, i.e., for $|B_g| = |P_1|$. 

\textit{Lines 15-34:} The \texttt{\textbf{ExpandA}} subroutine, at line 15, firstly computes $\bar{A}$, which requires to calculate the opposite remote set for (in the worst case) each nodes in $P_1$. This in turn exhibits a time complexity of $O(|P_1|*|P_2|)$. Then, \texttt{\textbf{ExpandA}} constructs $\bar{\texttt{A}}2\texttt{A}$ and $\texttt{B}2\texttt{A}$ by per-element checks for each $v_i \in \bar{\texttt{A}}$. In total, this requires  $O(|\bar{A}|*|\tilde{A}_g|*|P_2|)$ time, in the worst case, to compute $\bar{\texttt{A}}2\texttt{A}$. Similarly for $\texttt{B}2\texttt{A}$. The  check condition, at Line 8 within the subroutine  \texttt{\textbf{ExpandA}}, takes $O(|\texttt{B}2\texttt{A}(v_i)|*|P_2|)$. Overall, the construction of $\bar{\texttt{A}}2\texttt{A}$ and $\texttt{B}2\texttt{A}$ takes $O(|\bar{A}|*|P_1|*|P_2|) \subseteq O(|P_1|^2*|P_2|)$. 

After that, the procedures enters the \texttt{While} loop (Line 16-31 in Alg.~\ref{alg:01}), which exhibits two cases per iteration, namely, Case 1: $\texttt{A}(v_i) = \emptyset$ and Case 2: $\texttt{A}(v_i) \neq \emptyset$. 
\begin{itemize}
    \item Case 1 requires $O(\texttt{A})$ time for the vertex search.
    \item Case 2 computes the opposite remote set of a randomly selected vertex $v_i$ in $O(|P_2|)$ time, and computes the union of opposite remote set of $\texttt{A}(v_i)$, i.e., in $O(|\tilde{A}_g|*|P_2|)$ (in the worst case). If $|\bar{\texttt{A}}2\texttt{A}(v_i)|=1$, we need to spend  $O(|P_2|)$ time to calculate the opposite remote set of $v_j$.
\end{itemize}
Then, both the cases recompute $\texttt{A}, \bar{\texttt{A}}2\texttt{A}$ via \texttt{\textbf{ExpandA}} in $O(|P_1|^2*|P_2|)$ time. The \texttt{While} loop terminates when \texttt{A} is empty, which in the worst-case, requires $O(|P_1|)$ iterations. Accordingly to the above, we can state that the total \texttt{While} loop complexity is  $O(|P_1|^3*|P_2|)$.

The procedure then takes $O(|\hat{A}_g|*|P_2|)$ time to map each vertex in $\hat{A}_g$ to its opposite remote-set in $P_2$. As a result, the location of each extracted resource can be identified as $\{v_i, \overline{N}(v_i)\}$, where $v_i \in \hat{A}_g$.

Based on above analysis, the dominant term is $O(|P_1|^3*|P_2|)$. Thus, the time-complexity for determining the extractable volume $r^{\ell}_g(n)$ and the corresponding locations of the involved vertices in Alg.~\ref{alg:01} is polynomial.

\end{appendices}

\bibliographystyle{IEEEtran}
\bibliography{biblio.bib}

\begin{IEEEbiography}
[{\includegraphics[width=1in,height=1.25in,clip,keepaspectratio]{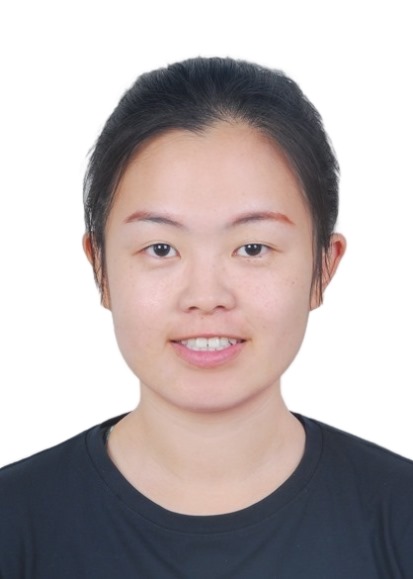}}]{Si-Yi Chen}\, received the Ph.D degree as 
an outstanding graduate in 2023 in School of Cyberspace Security from Beijing University of Posts and Telecommunications (China). Currently, she is a Postdoctoral fellow in University of Naples Federico II (Italy). She serves as a TPC member in IEEE International Conference on Quantum Computing and Engineering 2024 and 2025. Since 2022, she is a member of the Quantum Internet Research Group in University of Naples Federico II where she works on multiparty quantum networks, long-distance quantum communications and quantum entanglements.
\end{IEEEbiography}

\begin{IEEEbiography}
[{\includegraphics[width=1in,height=1.25in,clip,keepaspectratio]{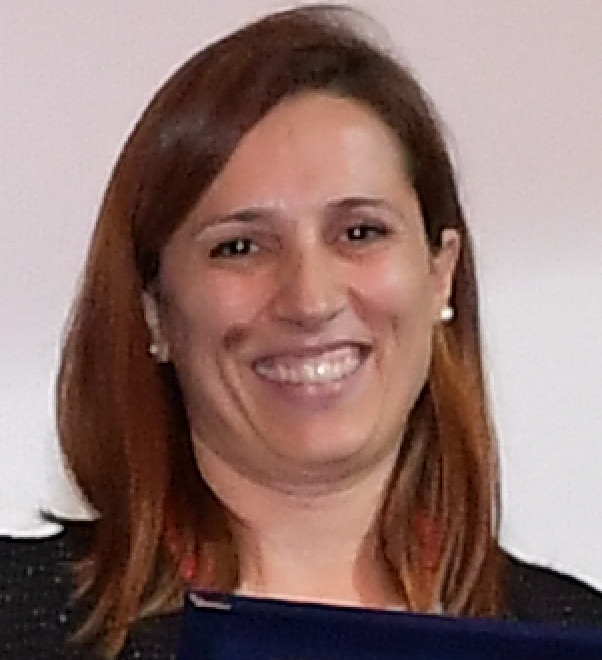}}]{Angela Sara Cacciapuoti} \, (Senior Member,
IEEE) is a Professor at the University of Naples Federico II, Italy, and co-founder of the Quantum Internet Research Group (www.quantuminternet.it). She is the PI of the ERC-Grant ``QNattyNet'', which aims to lay the foundations of a truly quantum-native Internet (www.qnattynet.quantuminternet.it). Prof. Cacciapuoti is a leading contributor to the theoretical and architectural foundations of quantum networking. Her work bridges quantum information theory, network design, and communication engineering, with a focus on enabling scalable quantum networks and hybrid quantum-classical architectures.
In recognition of her pioneering research, she has received several major international honors, including the ``2024 IEEE Communications Society Award for Advances in Communication'', the ``2022 IEEE ComSoc Best Tutorial Paper Award'', the ``2022 WICE Outstanding Achievement Award'', the ``2021 N2Women: Stars in Networking and Communications'', and the ``2023 IEEE ComSoc Distinguished Service Award (EMEA Region)''. She was also recently recognized as a \textit{Featured Author} on IEEE Xplore.
Prof. Cacciapuoti has been an IEEE ComSoc Distinguished Lecturer, delivering invited talks worldwide on Quantum Internet design and quantum network architectures. She currently serves as an Area Editor for IEEE Transactions on Communications and as a Senior Editor for the IEEE Journal on Selected Areas in Communications – Quantum Series. She is also on the editorial boards of npj Quantum Information, IEEE Transactions on Quantum Engineering, and IEEE Communications Surveys \& Tutorials. Previously, she served as Area Editor for IEEE Communications Letters (2019–2023), receiving the 2017 Exemplary Editor Award, and has held several IEEE leadership roles, including Vice-Chair and Publicity Chair of WICE and Treasurer of the IEEE Women in Engineering Affinity Group (Italy Section).
\end{IEEEbiography}

\begin{IEEEbiography}
[{\includegraphics[width=1in,height=1.25in,clip,keepaspectratio]{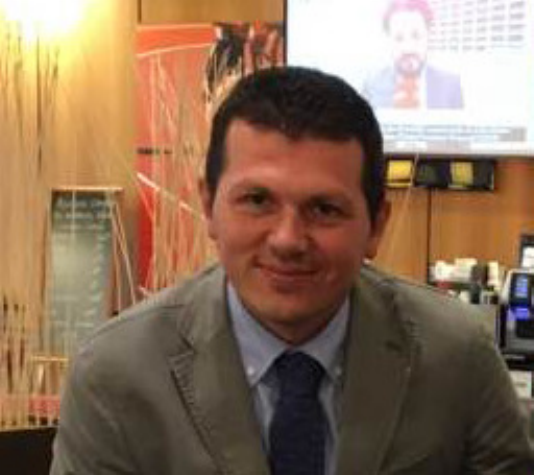}}]{Marcello Caleffi} \, (Senior Member, IEEE) is currently Professor of \textit{Advanced Quantum Networks} with the Department of Electrical Engineering and Information Technologies (DIETI), University of Naples Federico II, Naples, Italy, where he co-founded the Quantum Internet Research Group. His research has appeared in several premier IEEE Transactions and journals. He is the recipient of multiple awards, including the 2024 IEEE Communications Society Award for Advances in Communication and the 2022 IEEE Communications Society Best Tutorial Paper Award. He currently serves as Editor or Associate Editor for \textit{IEEE Transactions on Wireless Communications}, \textit{IEEE Transactions on Communications}, \textit{IEEE Transactions on Quantum Engineering}, \textit{IEEE Open Journal of the Communications Society}, and \textit{IEEE Internet Computing}. He has served as Chair and TPC Chair for several premier IEEE conferences. In 2017, he was appointed Distinguished Visitor Speaker by the IEEE Computer Society and was elected Treasurer of the IEEE ComSoc/VT Italy Chapter. In 2019, he was appointed as a member of the IEEE New Initiatives Committee by the IEEE Board of Directors, and in 2023, he was appointed as an IEEE ComSoc Distinguished Lecturer.
\end{IEEEbiography}

\end{document}